\documentclass{aa}  

\usepackage{graphicx}

\usepackage{subcaption}
\usepackage{txfonts}
\usepackage{color}
\def\zweb{WEBz\ }    
\def\zweb{PW-z\ }    %
\def\zw{$z_{web}\ $}             
\def\zcnn{$z_{CNN}\ $}
\begin{document} 

\title{{\it PhotoWeb redshift}: boosting photometric redshift accuracy with large spectroscopic surveys}
\titlerunning{\zweb: boosting photometric redshift accuracy}
\author{M. Shuntov \inst{1,2}
          \and J. Pasquet \inst{3}
          \and S. Arnouts \inst{1}
          \and O. Ilbert \inst{1}
          \and M. Treyer \inst{1},\\
                   E. Bertin \inst{2}
          \and S. de la Torre \inst{1}
          \and Y. Dubois \inst{2}
          \and D. Fouchez \inst{3}
          \and  K. Kraljic \inst{4}
          \and C. Laigle \inst{2}
          \and C. Pichon \inst{2,5}
          \and D. Vibert \inst{1}
          }

\institute{
Aix Marseille Universit\'e, CNRS, CNES, UMR 7326, Laboratoire d'Astrophysique de Marseille, Marseille, France 
 \and
Sorbonne Universit\'e, CNRS, UMR 7095, Institut d'Astrophysique de Paris, 98 bis bd Arago, 75014 Paris, France   \\
\email{shuntov@iap.fr}  
 \and 
 Aix-Marseille Universit\'e, CNRS/IN2P3, Centre de Physique des particules de Marseille, Marseille, France 
 \and
 Institute for Astronomy, University of Edinburgh, Royal Observatory, Blackford Hill, Edinburgh, EH9 3HJ, United Kingdom
 \and
  Korea Institute for Advanced Study (KIAS), 85 Hoegiro, Dongdaemun-gu, Seoul, 02455, Republic of Korea
          }
\date{Received on 20 December 2019 / Accepted on 6 March 2020}

\abstract{Improving  distance measurements in large imaging surveys is a major challenge to better reveal the distribution of galaxies on a large scale and to link  galaxy properties with their environments.
As recently shown,
photometric redshifts can be efficiently combined with the cosmic web extracted from  overlapping  spectroscopic surveys
to improve their accuracy. 
 In this paper we apply a similar method using
 a new generation of photometric redshifts based on a convolution neural network (CNN).   
 The CNN is trained on the SDSS images with the main galaxy sample (SDSS-MGS, $r\le 17.8$)  
 and the GAMA spectroscopic redshifts up to $r\sim 19.8$. The mapping of the cosmic web is obtained with 680,000 spectroscopic redshifts from the MGS and BOSS surveys. The redshift probability distribution functions (PDF), which are well calibrated 
 (unbiased and narrow, $\le$ 120 Mpc), intercept
 a few cosmic web structures along the line of sight.  Combining these PDFs with the density field distribution provides new photometric redshifts, \zw , whose accuracy is improved by a factor of two (i.e., $\sigma \sim 0.004(1+z)$) for galaxies  with $r\le 17.8$.  For half of them,  the distance accuracy is better than 10 cMpc. The narrower the original PDF, the larger the boost in accuracy. No gain is observed for original PDFs wider than 0.03. The final \zw PDFs also appear  well calibrated. The method performs slightly better for passive galaxies than star-forming ones, and for galaxies in massive groups since these populations better trace the underlying large-scale structure. Reducing the spectroscopic sampling by a factor of 8 still improves the photometric redshift accuracy by 25\%. Finally, extending the method to galaxies fainter than the MGS limit still improves the redshift estimates for 70\% of the galaxies, with a gain in accuracy of 20\%  at low z where the resolution of the cosmic web is the highest.
As two competing 
factors contribute  to the performance of the method, the photometric redshift accuracy and the resolution of the cosmic web, the benefit of  combining cosmological imaging surveys with spectroscopic surveys at higher redshift remains to be evaluated.
}
\maketitle


\section{Introduction}


%
%
Photometric redshifts are a key component for the exploitation of large imaging surveys \citep[see, e.g.,][for a review]{Salvato19}. They are a cheap alternative to spectroscopic surveys for the measurement of distances of millions of galaxies. They have been widely used to study the evolution of galaxy properties over cosmic time \citep[e.g.,][]{Ilbert2013, Madau2014, Davidzon17}
or to link galaxies with their dark matter halos \citep[e.g.,][]{Coupon2015},
and are essential to study the nature of dark energy. Weak lensing cosmological probes also need an accurate estimate of the mean redshift of the selected galaxy populations  \citep[]{Knox2006}, while the figure of merit of the baryon acoustic oscillation probe can  to some extent be improved by combining dense photometric samples with sparse spectroscopic surveys \citep[]{Patej2018}.  The derivation of robust redshift probability distribution functions (PDFs) is also necessary to understand the uncertainties attached to any of the above measurements \citep[]{Mandelbaum2008}. 

 The highly nonlinear mapping between the photometric space and the redshift space has been performed essentially via two broad techniques. The first,  
 template fitting   \citep[e.g.,][]{Arnouts1999,Benitez2000,Brammer2008}, matches the broadband photometry of each galaxy to the synthetic magnitudes of a suite of templates across a large redshift interval. This technique does not require a large spectroscopic sample for training, 
but it is often computationally intensive and  involves
poorly known parameters, such as dust attenuation, which can lead to degeneracies in color–redshift space. The second group of techniques includes   machine learning methods, such as artificial neural networks \citep[]{Collister2004}, k-nearest neighbors \citep[kNN,][]{Csabai2007}, self-organizing maps  \citep[SOM, ][]{Masters15,Davidzon19}, or random forest techniques \citep[]{Carliles2010}, which perform better within the limits of the training set \citep[]{Sanchez2014}, but the lack of spectroscopic coverage in some color--space regions, and at high redshift remains a major issue \citep[][]{Masters19}.  
For these reasons, 
 hybrid approaches have emerged 
 to optimize the photometric redshift PDF estimates \citep[e.g.,][]{CarrascoKind2014, Cavuoti2017, Hemmati19}. 

One limiting factor of these techniques is the information used as input. Magnitudes or colors are affected by choices of aperture size, PSF variations, and overlapping sources \citep[]{Hildebrandt2012}. 
In recent years the deep learning techniques, such as Convolutional Neural Networks (CNN), have bypassed  this limitation by dealing directly with multiband galaxy images at the pixel level, without relying on photometric feature extractions \citep[]{Hoyle2016, disanto2018, Pasquet2019}. As shown by \citet{Pasquet2019}, who trained a CNN on images from the SDSS spectroscopic sample, this method surpasses current machine learning photometric redshift estimates in the SDSS survey \citep[based on kNN,][]{Beck2017}. CNN photometric redshifts are almost free of bias with respect to  disk inclination and galactic reddening E$_{B-V}$, for example, while color-based photometric redshifts are not.  
 The associated PDFs are also well calibrated and provide a reliable indicator of the redshift uncertainty.  
 However, despite constant improvements in  photometric redshift techniques, even the best SED fitting (such as in the COSMOS field imaged in a large number of filters, \citealp{Laigle2016}) or deep learning methods \citep[]{Pasquet2019} hardly reach a redshift uncertainty $\sigma_z$ below $\sim 0.01$, which corresponds to a distance uncertainty of $\sim40$ cMpc (at $z\sim1$).

 Redshifts can also be predicted from the spatial distribution of galaxies on a large scale. The spatial cross-correlation between a photometrically selected sample and a reference sample with known spectroscopic redshifts offers an efficient way to infer the redshift distribution $N(z)$ of the photometric sample \citep[known  as the clustering redshift technique;  e.g.,][]{Matthews2010, Menard2013}. This method was extended with a hierarchical Bayesian model to simultaneously constrain $N(z)$ and the redshift of individual galaxies \citep[]{Leistedt2016, Sanchez2019}.

 With a similar methodology it is possible to improve the individual photo-z estimates by using the known
 galaxy density field, reconstructed from spectroscopic surveys \citep[]{Kovac2010, AragonCalvo2015} or the 3D tomography of the intergalactic medium with  neutral hydrogen absorption lines  \citep[at high redshift, ][]{Schmittfull2016, Lee2016}.
 The large-scale structure formation results from the anisotropic gravitational collapse of the primordial dark matter density fluctuations \citep[]{Zeldovich1970}, giving rise to large underdense regions bordered by sheet-like walls, which are framed by filaments connecting  density peaks (nodes). These features form the so-called cosmic web \citep[CW;][]{Bond1996},  identified in local galaxy surveys \citep[]{York2000,Colless2003}. Galaxies are  preferentially found 
in overdense regions of the underlying density field as a consequence of the biased formation of their dark matter halos
 \citep[]{Mo1996}. 
 The underdense regions appear almost empty of galaxies \citep[voids account for less than 5\% of luminous galaxies; ][]{AragonCalvo2015}, while they occupy almost 90\% of the volume of the universe \citep[]{AragonCalvo2010c, Cautun2014}. The vast majority of galaxies thus lie within the remaining 10\% composed of the dense regions distributed in a geometric pattern of walls, filaments, and nodes. The most massive galaxies live preferentially in the nodes (highest density regions), but segregation  also occurs in  filamentary regions where more massive or passive galaxies are closer to the center of the filaments \citep[][]{Malavasi2017,Kraljic2018}. 
 
 The galaxy density field can therefore provide strong priors on the location of a galaxy and help to  narrow its original photometric redshift PDF to a few more probable redshifts corresponding to the spikes of the intercepted density field along the line of sight, as proposed by \citet[][]{Kovac2010}.
 When anchored to the right density peak, the photometric redshift accuracy is increased up to the resolution of the reconstructed density field, usually a few cMpc, i.e., about 10 times better than current individual photometric redshift estimates.    
 This method has been further improved by \citet[][named the {\it PhotoWeb redshift} method]{AragonCalvo2015}.
 In addition to the density field, they introduced an extra term to mitigate the influence of the nodes (highest density peaks) in the resulting redshift PDF. For any point along the line of sight, this term scales inversely to the closest distance of any structure, allowing for a better contribution of less dense structures such as filaments and walls. They applied the {\it PhotoWeb reshift} to the SDSS sample. They reconstructed the cosmic web with the spectroscopic galaxies up to $z\sim0.12$ 
 and used the SDSS photometric redshifts of \citet{Csabai2007}, based on a k-NN method. Restricting the sample to galaxies with good photometric redshift accuracy ($\Delta z\le 0.015$), they showed that using the prior knowledge of the cosmic web yields photometric redshifts with Megaparsec accuracy.
 
 In the present paper, we adopt the same strategy as \citet[]{AragonCalvo2015}, but we push the analysis further in redshift and magnitude. 
 Our method, the  Photo Web
redshift technique 
(hereafter \zweb), relies on the cosmic web extractor DisPerSE \citep[]{Sousbie2011a}, and is applied to the CNN photometric redshifts from \citet[]{Pasquet2019}. 
 We explore the performance of the resulting photometric redshifts \zw as a function of  CNN PDF width, while making no pre-selection regarding their uncertainties, and we test the reliability of the final \zweb PDF. Furthermore, we analyze the performance of the method with respect to galaxy properties (galaxy types, group memberships) and how the resolution of the CW reconstruction impacts the \zw accuracy.
Our analysis also extends to galaxies two magnitudes fainter than SDSS using the GAMA survey.
 
 
 The outline of the paper is as follows. In section 2 we describe the photometric and spectroscopic dataset and show the \zcnn measurements. In section 3 we describe the \zweb method. The main results are presented in section 4, followed by the conclusion in section 5.       
 Throughout this paper we adopt a flat cosmology with  $\Omega_{\text{m}} = 0.307115 $ and the Hubble constant $H_0 = 67.77$ km s$^{-1}$ Mpc$^{-1}$.




%
\section{Spectroscopic and photometric redshift dataset}

\subsection{Spectroscopic redshift dataset}

\begin{figure*}[h]
\includegraphics[width=0.53\hsize, height=8cm]{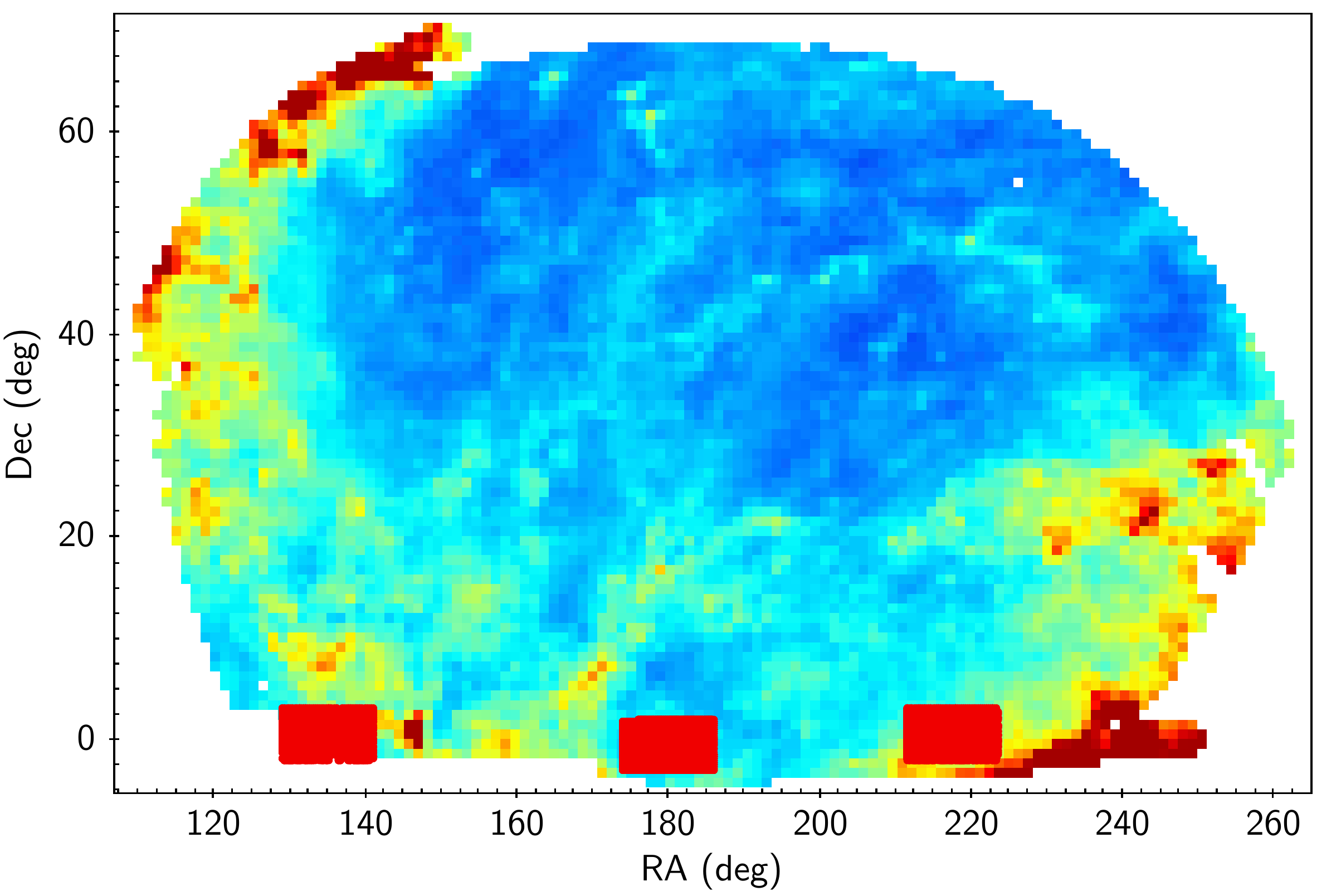}
\includegraphics[width=0.44\hsize,height=8.05cm]{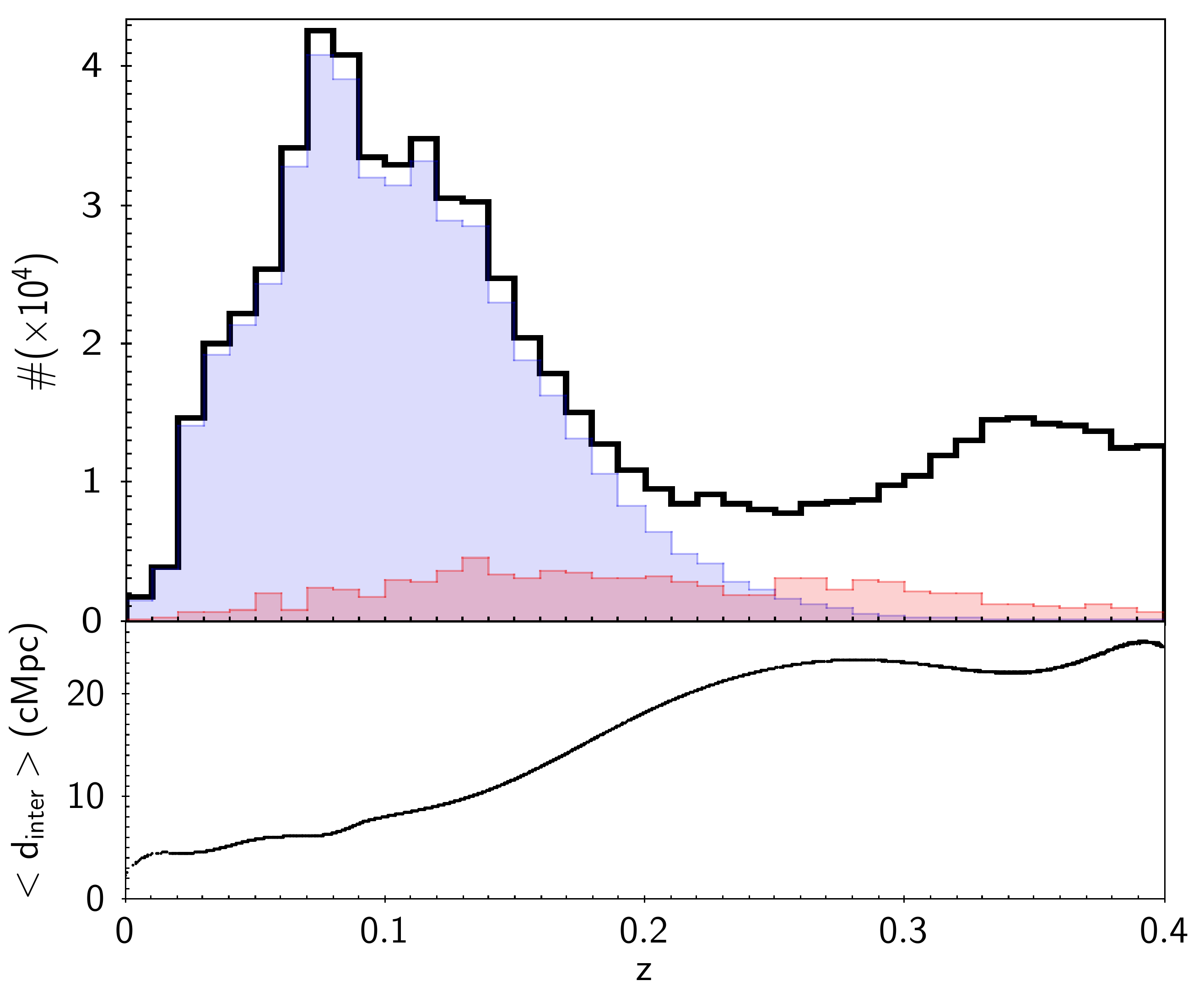}
\caption{Left: Footprints of the SDSS$+$BOSS sample, color-coded with the galactic reddening excess ($E(B-V)$), and the three equatorial GAMA fields (large red rectangles). Right top panel: Redshift distributions of the SDSS MGS sample ($r\le 17.8$, filled light blue histogram), the GAMA sample ($r\le 19.0-19.8$, filled light red histogram), and the whole spectroscopic sample (MGS$+$LRGs$+$BOSS samples; solid black line) used to reconstruct the cosmic web. Right bottom panel:  Mean intergalactic comoving distance of the whole spectroscopic sample (see text).
}
\label{fig:sample}
\end{figure*}

To perform this analysis, we use the SDSS and BOSS spectroscopic samples from the data release 12 \citep[DR12, ][]{Alam2015} and the GAMA spectroscopic samples from the data release 3 \citep{Baldry2018}.
The characteristics of each sample are as follows:
\begin{itemize}
\item We use the main galaxy sample of the SDSS (hereafter MGS) to train and validate our photometric redshift estimates. It is limited to galaxies with dereddened Petrosian magnitudes $r\le 17.8$. We only use the large contiguous region shown in Fig.~\ref{fig:sample} (left panel), covering $\sim$7400 deg$^2$ and containing  $\sim$ 480,000  galaxies.
The redshift distribution is shown in Fig.~\ref{fig:sample} (right panel). 
\item  The spectroscopic sample used to reconstruct the CW consists of the MGS sample completed by the luminous red galaxy sample (LRG) and by the BOSS sample for a total of $\sim$ 686,000 galaxies up to $z=0.4$. This redshift limit encompasses all the MGS galaxies. The redshift distribution is shown in Fig.~\ref{fig:sample} (right panel).
 In the bottom panel, we show the evolution of the spatial density as a function of redshift, characterized by the mean intergalactic comoving distance\footnote{ $<\hbox{d}_{\rm inter}>= 1/\sqrt[3]{\varphi(z)}$, where $\varphi(z)$ is the selection function taking into account the density variation with the radial distance induced by the flux limits and color selections of the different samples.}.
Between $z\sim 0.15$ and $z\sim 0.25$, the target density becomes sparse, providing a coarse representation of the CW above $z\sim 0.2$. 
\item We also derived a second set of photometric redshifts trained with the GAMA spectroscopic survey which is two magnitudes deeper than the MGS sample. It consists of two fields with spectroscopic redshifts down to $r= 19.0$ (namely G09 and G12) and one field down to $r = 19.8$ (G15). These three fields cover 180 deg$^2$ and overlap the SDSS-BOSS footprint, as shown in Fig.~\ref{fig:sample} (left panel). As suggested by the GAMA team, we restrict the sample to the most secure redshifts (nQ $\ge 3$), namely $\sim$99,500 galaxies. The total redshift distribution is shown in Fig.~\ref{fig:sample} (right panel). In the deepest field (G15), 4\% of the galaxies are located at $z>0.4$ and will be ignored in the rest of the paper.
\end{itemize}
%
%
\subsection{Photometric redshifts with SDSS-MGS}
\label{subsec:photozMGS}
Our first set of photometric redshifts is trained and validated with the SDSS-MGS sample \citep{Pasquet2019}. They are estimated with a convolutional neural network (CNN), which is a special type of multilayered neural network. The input data are 64x64 pixel images centered on the galaxy coordinates in the five bands of the SDSS imaging survey ({\it ugriz}). The architecture of the CNN is detailed in \citet[]{Pasquet2019}. In brief, it is composed of several convolution and pooling layers followed by fully connected layers. The convolution part of the network is organized in multi-scale blocks called inception modules to treat the signal at different resolution scales \citep{Szegedy2015}. The redshift values are estimated as a classification problem, where each class corresponds to a narrow redshift bin $\delta z$ (here 180 redshift bins between $0\le z\le 0.4$). The network assigns a probability to each redshift bin, which is used as a probability distribution function (PDF). We define the redshift value as the weighted mean of the PDF ($z_{CNN}=\sum_{k} z_{k} PDF_k$). The power of this technique relies in the exploitation of all the information available in the images at the pixel level, without any 
prior
feature extraction. 
 %

To assess the performance of the method, we adopt the same statistics used by \citet{Pasquet2019}:
\begin{itemize}
\item the \textbf{residuals}, $\Delta z = (z_{\text{CNN}}-z_{\text{spec}})/(1+z_{\text{spec}})$;
\item the \textbf{bias}, $< \Delta z >$, defined as the mean of the residuals;
\item the \textbf{MAD deviation} (Median Absolute Deviation)\footnote{Strictly speaking, this is the standard deviation $\sigma$ estimated from the MAD deviation for normally distributed data: $\sigma\approx 1.4826\times \text{MAD}$.}, defined as $\sigma_{\text{MAD}} = 1.4826 \times \text{median}(|\Delta z - \text{median}(\Delta z)|)$.
\item the fraction of \textbf{outliers}, $\eta$, defined as the fraction of galaxies with $|\Delta z|>0.05$;
\end{itemize}

\begin{figure}[ht]
\includegraphics[width=0.95\hsize,height=6cm]{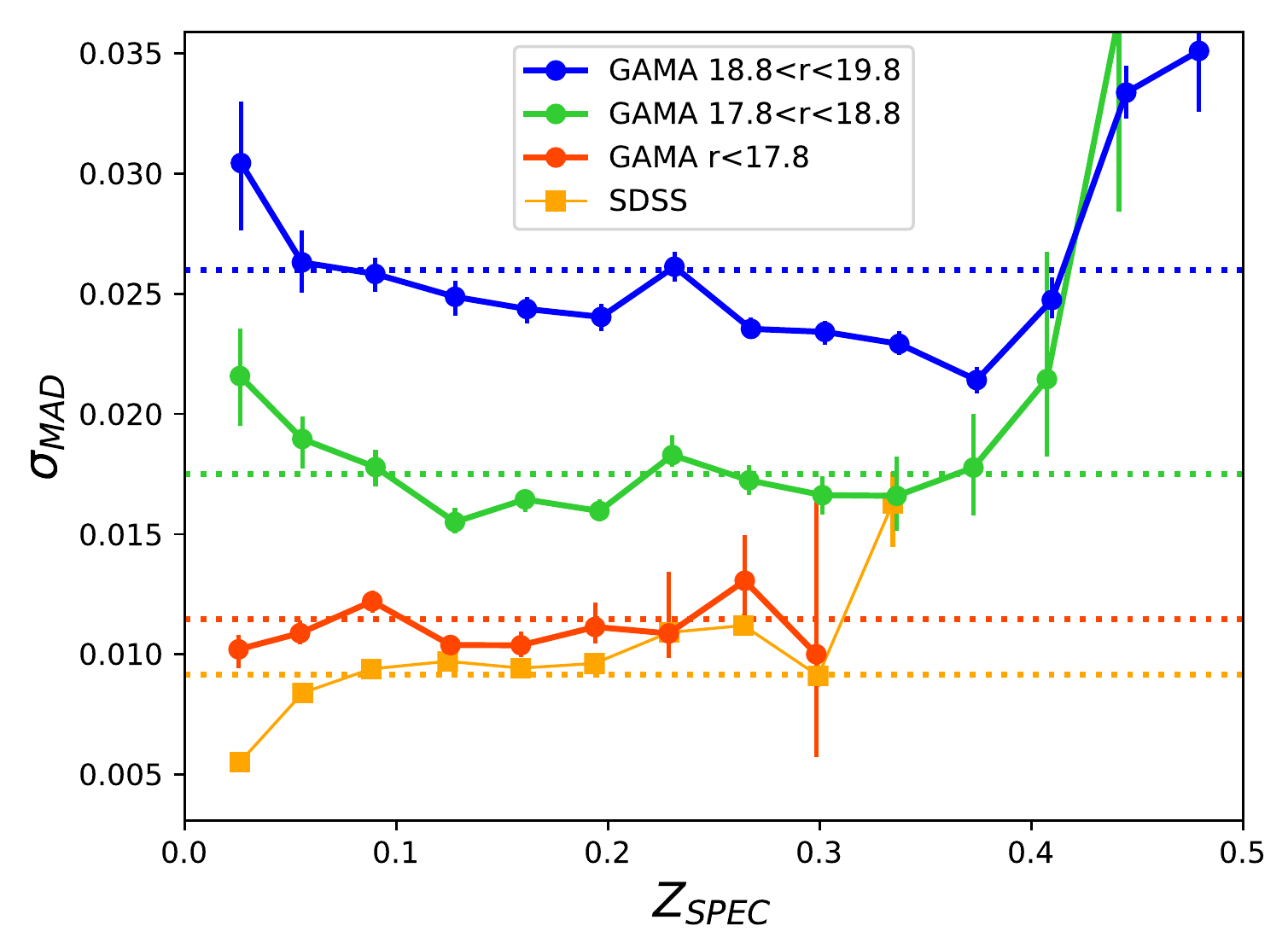}
\caption{Accuracy of the photometric redshift point estimates ($\sigma_{\text{MAD}}$) for the SDSS-MGS and GAMA surveys as a function of spectroscopic redshift (solid lines) and 
the complete subsamples (dotted lines).
}
\label{fig:sig}
\end{figure}

The CNN photometric redshifts are highly accurate at the depth of the SDSS-MGS sample ($r\le 17.8$), with $\sigma_{\text{MAD}}$ lower than 0.01. In Fig.~\ref{fig:sig} we show the evolution of $\sigma_{\text{MAD}}$ as a function of redshift. The behavior of the photometric redshift accuracy is relatively independent of redshift up to $z\sim 0.3$ for the MGS, with less than 2\% of the MGS being above this redshift 
%
%

 Of particular interest is the reliability of the redshift PDF derived by the CNN. To evaluate the predictive power of the PDFs we use the probability integral transform statistic \citep[PIT;][]{Polsterer2016, Pasquet2019}. For each galaxy the PIT is measured as the cumulative PDF (CDF) up to the spectroscopic redshift, $z_s$ $\left( {\rm CDF}(z_s) =\int_{0}^{z_s} PDF(z)\ dz \right)$.
A flat distribution of the PIT values in a given sample indicates that the PDFs are not biased with respect to the spectroscopic redshifts. They are neither too narrow nor too wide, whereas convex or concave distributions point to under- or  overdispersed PDFs, respectively \citep{Polsterer2016}. 
A negative or positive slope in the PIT distribution  indicates a systematic bias (over- or underpredicted redshifts, respectively).   
We find a nearly flat PIT distribution except at the extreme values that are slightly underpopulated, which suggests that the PDFs are marginally too broad \citep[see Fig.10 in][]{Pasquet2019}.

 \subsection{Photometric redshifts with GAMA}
\label{subsec:photozGAMA}

We also created a second set of photometric redshifts using GAMA as the training sample, which is two magnitudes deeper than the MGS. 
The characteristics of the input images remain the same as described in Sect. \ref{subsec:photozMGS} (64x64 pixel images from the SDSS imaging survey in five bands {\it ugriz}).

 As the size of this training database is smaller than the MGS sample and extends to higher redshift, we had to adapt the architecture. 
  The new architecture is shallower  and alternates five convolution layers and three pooling layers, followed by two fully connected layers. The classifier consists of 300 bins between $0\le z\le 0.6$. The total number of parameters (9,433,196) is reduced compared to the CNN trained on the MGS sample, in order to avoid overfitting. A clear difficulty is the low signal-to-noise ratio of the SDSS images for the GAMA sources  fainter than the MGS (up to 2 magnitudes) that degrades the performance. To tackle this problem, we optimized the choice of the activation functions and the pooling operations. We used the hyperbolic tangent as an activation function of the first layer, 
 which saturates the signal at high values, thus narrowing its range in order to facilitate the learning stage. 
 Then we used max 
pooling instead of average pooling operations in order to give more weight to the flux of the galaxy than to the noise.
 

Figure \ref{fig:sig} shows the  CNN redshift precision as a function of  spectroscopic redshift  for the GAMA training. The lower accuracy obtained for the GAMA sources at bright magnitude ($r<17.8$) compared to the SDSS-MGS training is due to the smaller size of the training set and the simpler CNN architecture, but it is comparable.  
At fainter magnitudes, $\sigma_{\text{MAD}}$ gradually increases as a result of the decreasing S/N 
in the five photometric bands, in particular in the $u$ and $z$ bands where the majority of galaxies with $r\ge19.5$ have a S/N lower than 10. As in \citet[]{Pasquet2019}, we compare our results with the photometric redshifts of \citet[]{Beck2017} available in SDSS DR12 and estimated with a k$-$NN method \citep[local linear regression,][]{Csabai2007}. As for the SDSS-MGS, the CNN redshifts performs better, with a MAD deviation $\sigma_{\rm MAD}=0.017$ and 0.026 in the two faint magnitude bins compared to $\sigma_{\rm MAD}=$0.022 and 0.034 for the k$-$NN method. 

\begin{figure}[ht]
\includegraphics[width=0.95\hsize,height=6cm]{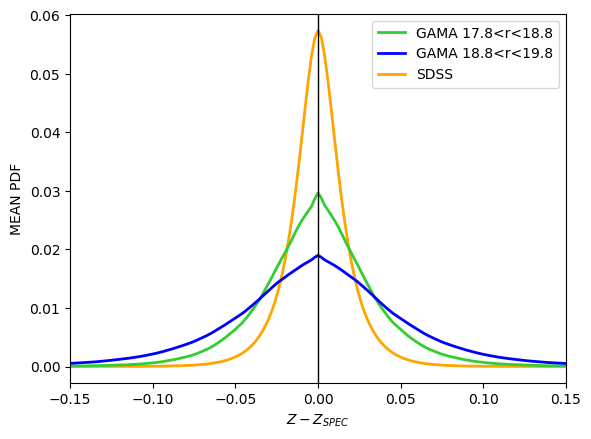}
\caption{Mean PDFs, recentered on the individual spectroscopic redshifts, for the SDSS and GAMA surveys. We define three subsamples according to their magnitude range: $r<17.8$ (orange line), $17.8<r<18.8$ (green line), and $18.8<r<19.8$ (blue line).}
\label{fig:PDF}
\end{figure}

In 
Fig.~\ref{fig:PDF}, we show the mean PDFs recentered at the spectroscopic values for the same samples. That of the SDSS-MGS sample is significantly narrower than the mean PDF of the GAMA sample, especially at faint magnitude. This broadening of the PDFs with magnitude, in line with the increase in $\sigma_{\rm MAD}$ at lower S/N, reflects the increasing uncertainty on the photometric redshifts in a reliable way since we also find the PIT distribution to be equally flat at all magnitudes.  
 
Finally, we note that the behavior of the photometric redshift accuracy is relatively independent of redshift (Fig.~\ref{fig:sig}) up to $z\sim 0.4$ for GAMA in both magnitude intervals. In the following we restrict our analysis to $z<0.4$. 
%

In conclusion, the CNN method provides photometric redshifts  that are accurate for  both the MGS and the GAMA samples and unbiased up to $z\sim 0.3$ and $z\sim 0.4,$ respectively. 
The distance accuracy\footnote{The distance uncertainty can be expressed as $\delta D=[c(1+z)/H(z)].[\delta z/(1+z)]$  \citep[]{Schmittfull2016}, where $H(z)=H_0.\sqrt{(1+\frac{\Omega_M}{\Omega_{\Lambda}}(1+z)^3)/(1+\frac{\Omega_M}{\Omega_{\Lambda}})}$. This leads to $\delta D$(Mpc)=4500\ $\sigma_{\rm MAD}$ at $z\sim 0.1$.}
 is 40 - 112 Mpc at $<z>=0.10$ for $\sigma$=0.009 (SDSS) and 0.025 (GAMA), respectively, which correspond to the typical size of the largest void. In the following section we investigate whether the combination of the PDFs from the different samples and the knowledge of the cosmic web  environment reconstructed with the spectroscopic surveys can further improve the photometric redshift estimates.



%
\section{The \zweb method}
\subsection{Method} \label{method}
\begin{figure}
\centering
\includegraphics[width=0.99\hsize]{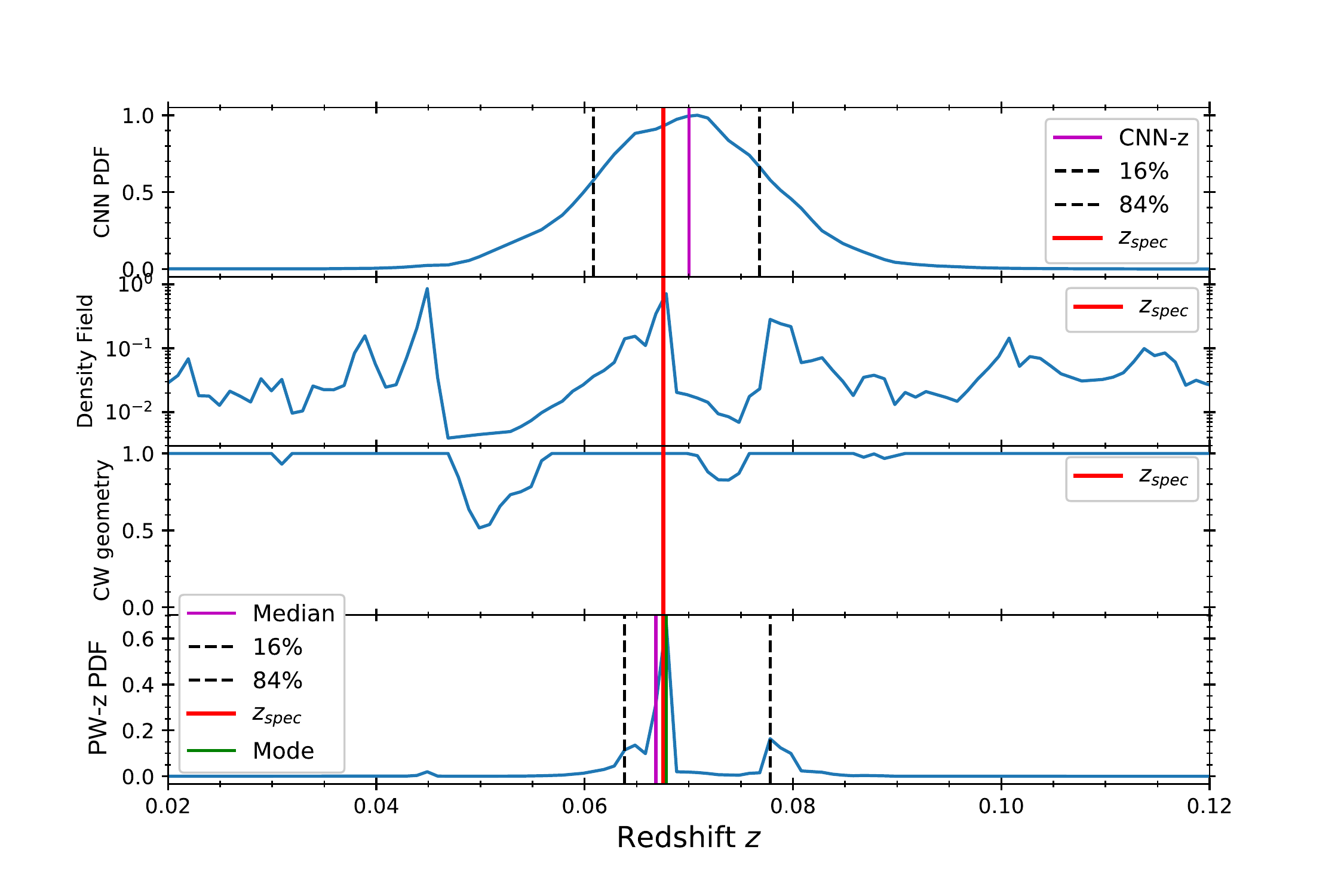}     
\caption{Illustration of the \zweb technique. From top to bottom: \textbf{i)}  Initial CNN photometric redshift PDF, \textbf{ii)}  CW density field, 
\textbf{iii)}  Probability taking into account the closest distance of any geometric structure of the CW (Eq. \ref{eq: PDF-CW}),
\textbf{iv)}  Final \zweb PDF. 
The red line indicates the $z_{\text{spec}}$, the magenta line  the $z_{CNN}$ (top panel) and the median of the \zweb PDF (bottom panel), the green line indicates the mode of the \zweb PDF, and the dashed lines the $68\%$ confidence interval.
}
\label{fig:zweb}
\end{figure}

 As described in \citet{AragonCalvo2015}, the main idea of the \zweb technique 
  is to exploit the galaxy distribution of a spectroscopic survey in order to improve the photometric redshift of other galaxies that are expected to be embedded in this distribution. The broad PDFs derived from a given photometric redshift technique (here the CNN-based $PDF_{\text{CNN}}(z)$) are combined with the probability distribution function of the density field ($P_{\text{den}}(z)$, reconstructed from the spectroscopic survey) along the line of sight (LoS) as follows: 
\begin{equation} \label{P_pw}
PDF_{\text{PW-z}}(z) = PDF_{\text{CNN}}(z) \ .\  P_{\text{dens}}(z)\ .\ P_{\text{CW}}(z) .\end{equation}
 Figure \ref{fig:zweb} illustrates the method. The original PDF of the galaxy derived from the CNN is shown in the top panel  with its mean redshift estimate and uncertainties (68\% confidence interval), as well as the spectroscopic redshift. The reconstructed density field, illustrated in the second panel, shows the crossing of several structures along the LoS (alternate low- and high-density regions, spanning a wide dynamical range). To prevent the final PDF to be anchored on the densest group or cluster regardless of the vicinity of less dense structures (filaments or walls), \citet{AragonCalvo2015} introduced an additional term taking into account the geometry of the CW, beyond the density. 
 At each redshift along the LoS, the shortest distance to any of the CW features (walls, filaments, nodes) is estimated and converted into a probability
 $P_{\text{CW}}(z)$ \footnote{This is our own parameterization since it is not explicitly described in \citet[]{AragonCalvo2015}. We choose this function empirically, having in mind the geometry of the reconstructed CW; we tested several versions of this function and found that the exact values do not significantly impact the results.} 
as follows:
 \begin{equation}
 \label{eq: PDF-CW}
    P_{\text{CW}}(z) =
  \begin{cases}
     1       & \quad \text{if } d_{\text{ns}} \leq 10\\
    \dfrac{(10-d_{\text{ns}})}{20}+1      & \quad \text{if } 10 < d_{\text{ns}} < 30\\
    0       & \quad \text{if } d_{\text{ns}} \geq 30
  \end{cases}
 ,\end{equation}
 where $d_{\text{ns}}$ is the 3D Euclidean distance to the nearest CW structure in cMpc.
 This is illustrated in the third panel. This term alleviates the dominating influence of neighboring nodes on which filaments connect.
 The density field and CNN PDFs are resampled with $\delta z = 10^{-3}$ in $0<z<0.4$ with a linear interpolation, and the $P_{\text{CW}}(z)$ is also computed at these points. This results in final $PDF_{\text{PW-z}}$ with the same sampling. 
 The resulting PDF ($PDF_{\text{PW-z}}(z)$) is shown in the bottom panel with its median, mode, and 68\% confidence interval. In that specific case, the original PDF is shrunk around the highest density peak, which happens to correspond to the spectroscopic redshift of this galaxy. Other illustrations of \zweb PDFs are shown in Appendix A. 
\subsection{Density field and CW reconstruction}

 To extract the density field from the spectroscopic redshift survey and reconstruct the CW with the complex connectivity of its different components (nodes, filaments, and walls), we use the Discrete Persistent Structure Extractor code \citep[DisPerSE;][]{Sousbie2011a}, a geometric 3D robust ridge extractor working directly with the discrete 3D data points. 
 As demonstrated in \citet{Sousbie2011b}, DisPerSE can identify fairly poorly sampled structures, which will prove critical in what follows. 
 
 The underlying density field is computed from the discrete distribution of galaxies using the Delaunay Tessellation Field Estimator (DTFE) technique \citep{Schaap2000}. The DTFE is used to generate a simplicial complex, i.e., a geometric complex of cells, faces, edges, and vertices mapping the whole volume. The value of the density field, $f$, is estimated at each vertex of this complex and scales with the inverse of the volume of each tetrahedron. It naturally maps the anisotropic distribution of galaxies and can be linearly interpolated at any position of the volume, and within holes (unobserved or masked regions) in the spectroscopic survey \citep[see the example in][]{AragonCalvo2015, Malavasi2017}. 
 In Eq.~1 we use the density contrast, defined as $1+\delta=f/\varphi(z)$, where the local density, $f$, is normalized by the mean density, $\varphi(z)$, which decreases with radial distance. Along each LoS, $P_{\text{dens}}(z)$ is normalized to unity. 

 To identify the topological structures of the CW (nodes, filaments, and walls), DisPerSE relies on discrete Morse and persistence theories. Morse theory provides a framework in which to extract from $f$ the critical points where the discrete gradient, $\nabla f$, vanishes (e.g., maxima, minima, and saddle points). It then connects  critical points via the field lines tangent to $\nabla f$ in every point, while relying on a geometrical segmentation of space, known as the discrete Morse complex, within which all the field lines have the same origin and destination. This segmentation defines distinct regions called ascending and descending manifolds. 
 The morphological components of the CW are then identified from these manifolds\footnote{Ascending 3-manifolds trace the voids, ascending 2-manifolds trace the walls, ascending 1-manifolds trace the filaments, with their end points connected  onto the maxima (the peaks of the density field).}. The finite sampling of the density field introduces noise to the detection of structural features. DisPerSE makes use of persistent homology  to pair the critical points according to the birth and death of the relevant feature.  The ``persistence'' of a feature  is assessed by the relative density contrast of the density of the critical pair chosen to pass a certain signal-to-noise threshold. The noise level is defined relative to the variance of persistence values obtained from random sets of points and estimated for each type of critical pair. 
 This thresholding eliminates critical pairs and simplifies the corresponding discrete Morse complex, retaining only its most significant features. 
 
 By construction this method is scale invariant and builds a network which adapts naturally to the uneven sampling of observed catalogues. To prevent the spurious detection near the edges of the survey, DisPerSE encloses each field into a larger volume. New particles are added by extrapolating the density field measured at the boundary of the survey \citep[see, e.g.,][for illustrations]{Sousbie2011a, Kraljic2018}. Once the different manifolds are attached to specific topological features, we can estimate the closest structure (node, filament, or wall) from any point along a specific line of sight to derive the associated probability ($P_{\text{CW}}(z)$). As in  \citet[]{AragonCalvo2015}, we  find that this additional term is a second-order correction 
 and only affects the results described below at a subpercent level.
 Finally, we do not correct for the Finger-of-God effect, in contrast to what is done in \citet[]{AragonCalvo2015}. The large redshift range considered in this work (0<z<0.4) introduces a variable sampling of the density field, preventing us from performing an efficient group reconstruction at all z. The ``isotropizing'' of the groups also introduces an uncertainty in the redshift assignment. This may prevent us from getting highly accurate redshift at sub-Mpc scales for some galaxies, but as discussed later it still provides a significant improvement with respect to the original photometric redshifts.      
%
\subsection{Adopted strategy}
\label{metrics}
 The density field is estimated from the combined SDSS (MGS and LRG) and BOSS samples up to $z\sim$0.4 and the CW features are reconstructed with a 3$\sigma$ persistence threshold.\\
 To test the performance of the \zweb method, we select a sample of galaxies that were neither used in the CNN training nor used in the CW reconstruction. 
 In practice, we randomly select $\sim$17,000 galaxies from each test sample of the cross-validations created by \citet[]{Pasquet2019}, and reconstruct the CW with all the remaining galaxies. In this way the small fraction ($\sim$2\%) of  galaxies removed has no impact on the CW reconstruction, thus on the results of the \zweb method. We repeat this operation five times for each of the five  cross-validations.      
We find that the results are consistent throughout the five subsamples.
In the next sections, the 85,000 test galaxies are used to measure the performances of the \zweb method.
%
%




%
\section{Results}
\begin{figure*}
\centering
\includegraphics[width=\hsize]{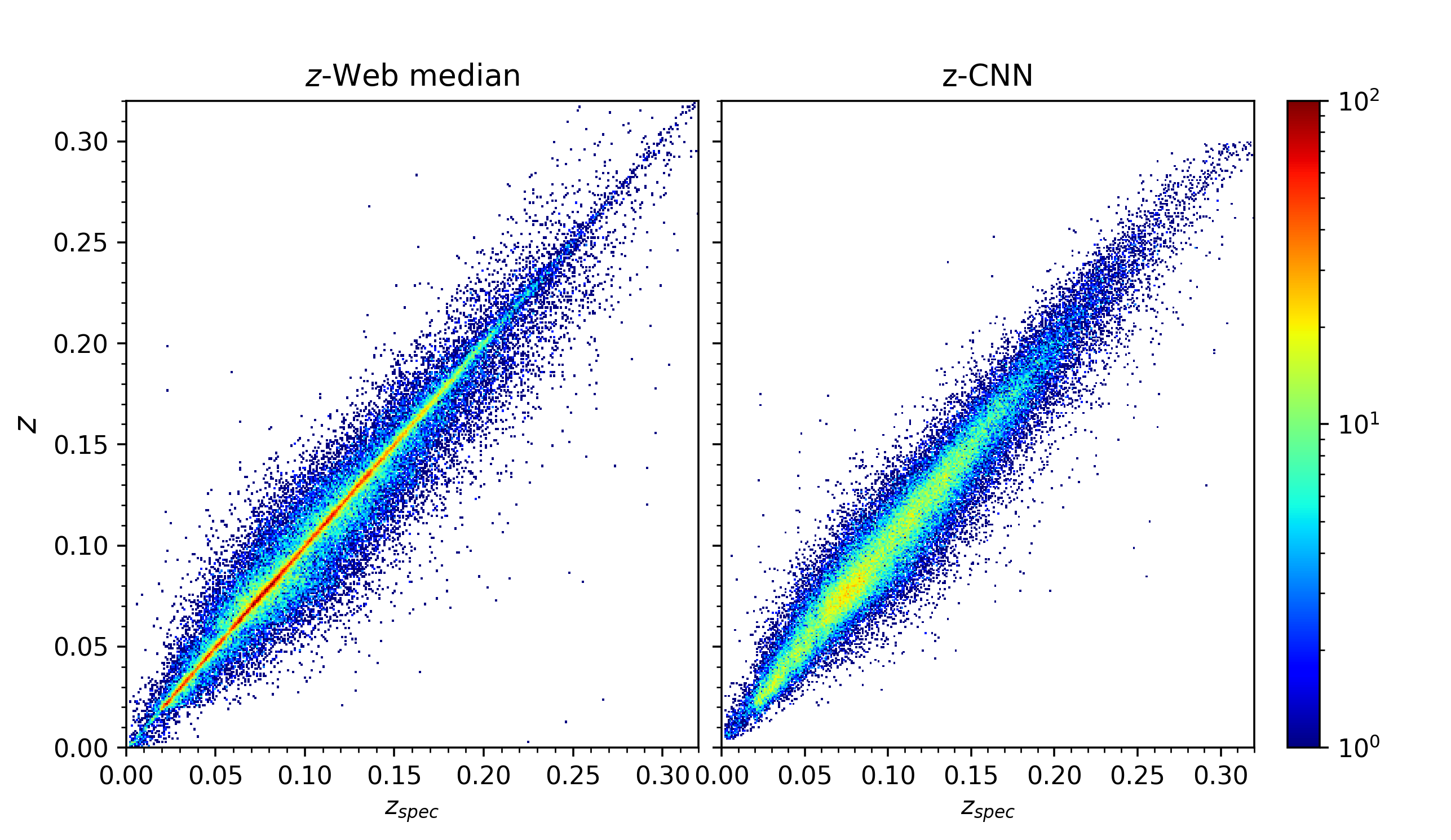} 
\caption{Comparison between the photometric and spectroscopic redshifts. Left: \zw  defined as the median of the \zweb PDF. Right: original \zcnn.} 
\label{fig:z-scatt}
\end{figure*}
The \zw redshifts are obtained from the final \zweb PDF derived with Eq.~1. While in \citet[]{Pasquet2019} we adopted the mean value of the PDF as point estimate, \zcnn, in the following we   consider different definitions for the \zweb redshift, \zw, with the mode, the mean, and the median of the \zweb PDF, and we   explore their relative performance. 
The mode anchors the \zw to the strongest density peak and as such best illustrates the method. The mean and median rely on the full \zweb PDF while still benefiting from the narrowing of the original CNN PDF.       

\begin{table}
\caption{Performance of the different \zw and \zcnn redshift estimates. 
} 
\label{tab:trials}      
\centering          
\begin{tabular}{| c | c | c | c | c |} 
\hline       
Selection & $\sigma_{\text{MAD}}$&$\eta$& \multicolumn{2}{|c|}{$\Delta$\zw$<$} \\
          &   ($\sigma_S$)     &      &  10cMpc  & $\Delta z_{CNN}$\\
          &  $\times10^{-3}$   &(\%)  & (\%)              & (\%)    \\ \hline \hline
\multicolumn{5}{|c|}{ \zw  (mode) } \\ \hline 
   full sample   & 3.8 (0.6) & 0.83 & 48 & 77 \\   \hline
  width $<$ 0.03 & 2.9 (0.7) & 0.13 & 52 & 79 \\    \hline
   width $<$ 0.02 & 2.1 (0.6)& 0.02 & 58 & 83 \\  \hline \hline
\multicolumn{5}{|c|}{ \zw  (median) } \\ \hline 
   full sample    & 4.5 (1.2)& 0.44 & 44 & 88 \\  \hline
   width $<$ 0.03 & 3.5 (1.2)& 0.07 & 49 & 89 \\  \hline
   width $<$ 0.02 & 2.6 (1.1)& 0.02 & 55 & 92 \\ \hline  \hline
\multicolumn{5}{|c|}{ \zw  (mean) } \\ \hline 
   full sample    & 6.6 (2.2)& 0.31 & 31 & 97 \\   \hline
   width $<$ 0.03 & 5.4 (1.9)& 0.04 & 36 & 97 \\  \hline
   width $<$ 0.02 & 4.0 (1.5)& 0.02 & 45 & 98 \\   \hline  \hline
\multicolumn{5}{|c|}{ \zcnn } \\ \hline
 full sample    &  9.2 (5.9) & 0.28 & 21 & $-$ \\  \hline
 width $<$ 0.03 & 7.9  (5.3) & 0.04 & 24 & $-$ \\  \hline
 width $<$ 0.02 & 6.3  (4.8) & 0.02 & 29 & $-$ \\ \hline 
\end{tabular}
 \tablefoot{Performance of the different \zw (top three blocks) and \zcnn (bottom block) redshift estimates ($\sigma_\text{MAD}$, $\eta$) for the whole sample (85,000 galaxies) and two subsets with CNN PDFs widths $\le$0.03  and 0.02 (corresponding to 75\% and 40\% of the whole sample, respectively). In the second column the $\sigma_S$ value from the double-Gaussian modeled residual function is also reported (see text). The last two columns show the fraction of galaxies with residuals $\Delta z \le$10 cMpc and with \zw accuracy better than the \zcnn.}
\end{table}
%
\subsection{Global performance of the \zweb method}
\begin{figure}
\centering
\includegraphics[width=1.1\hsize]{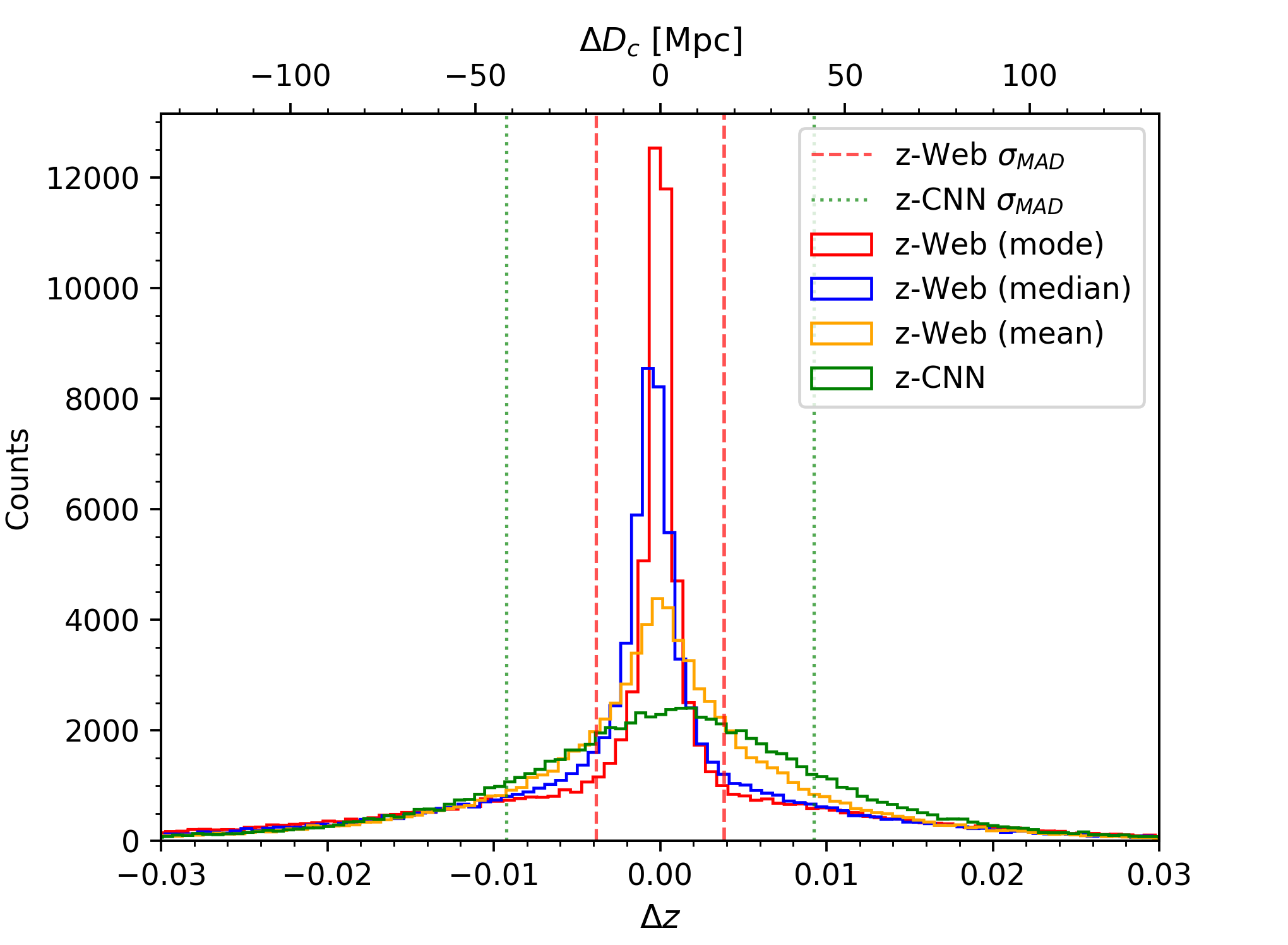}
\includegraphics[width=1.0\hsize,]{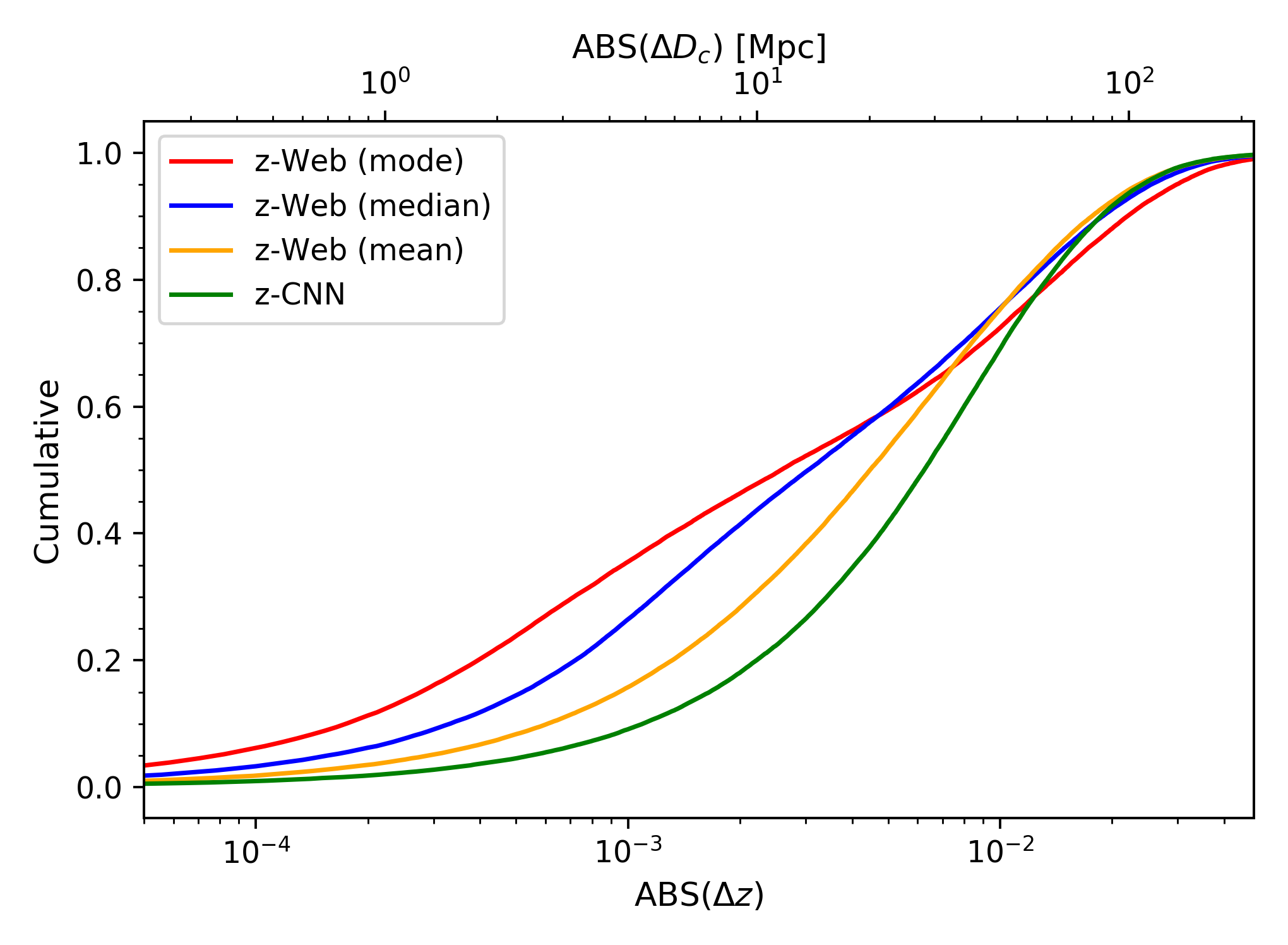}
\caption{Differential (top) and cumulative (bottom) histograms of the residuals for the \zw (mode: red, median: blue, mean: orange) and \zcnn (green). The dashed and dotted vertical lines indicate the respective $\sigma_{\text{MAD}}$. The distance uncertainties in comoving Mpc are shown on the top axis (assuming $<z>\sim0.1$).}
\label{fig:z-err}
\end{figure}

 Figure~\ref{fig:z-scatt} compares the \zw and \zcnn redshifts with the spectroscopic redshifts for the full sample. The \zw redshifts are significantly improved compared to the \zcnn, with an increased fraction of sources along the identity line while a modest increase in catastrophic failures is observed.   
  This is quantified in Table~\ref{tab:trials}, and is illustrated in Figure~\ref{fig:z-err} for the three definitions of \zw.
  Figure~\ref{fig:z-err} (top panel) shows the boost of highly accurate \zw redshifts from a factor of 2 (for \zw mean) to a factor of 6 (for \zw mode) when considering the full test sample. 
 The $\sigma_\text{MAD}$ is reduced by a factor of $\sim$ 2.5, 2.0, and 1.4 for \zw based on the mode, median, and mean, respectively (Table~\ref{tab:trials}). This translates into a gain in distance uncertainty from $\sim 40$ cMpc (for \zcnn) down to $\sim17$ cMpc (for \zw mode). 
 About half of the sample (45\%) has a redshift accuracy better than 10cMpc (for \zw  mode and median), 
  more than twice the fraction for \zcnn (20\%; see Fig.~\ref{fig:z-err}, bottom panel). \\ 
  As a drawback of the \zweb method, about 25\% of \zw based on the mode have worse estimates than \zcnn (Table~\ref{tab:trials} and Fig.~\ref{fig:z-err}, bottom panel). This happens when the galaxy is associated with the wrong structure of the density field and mainly impacts the \zw based on the mode of the PDF where the fraction of less accurate redshifts than \zcnn becomes significant. By adopting the \zw based on the median instead of the mode, we can mitigate this bias and reduce the number of galaxies with worse redshifts than \zcnn to $\sim$10\%, while keeping a high fraction of galaxies with significantly improved redshifts.

\begin{figure}
\centering
\includegraphics[width=1.1\hsize]{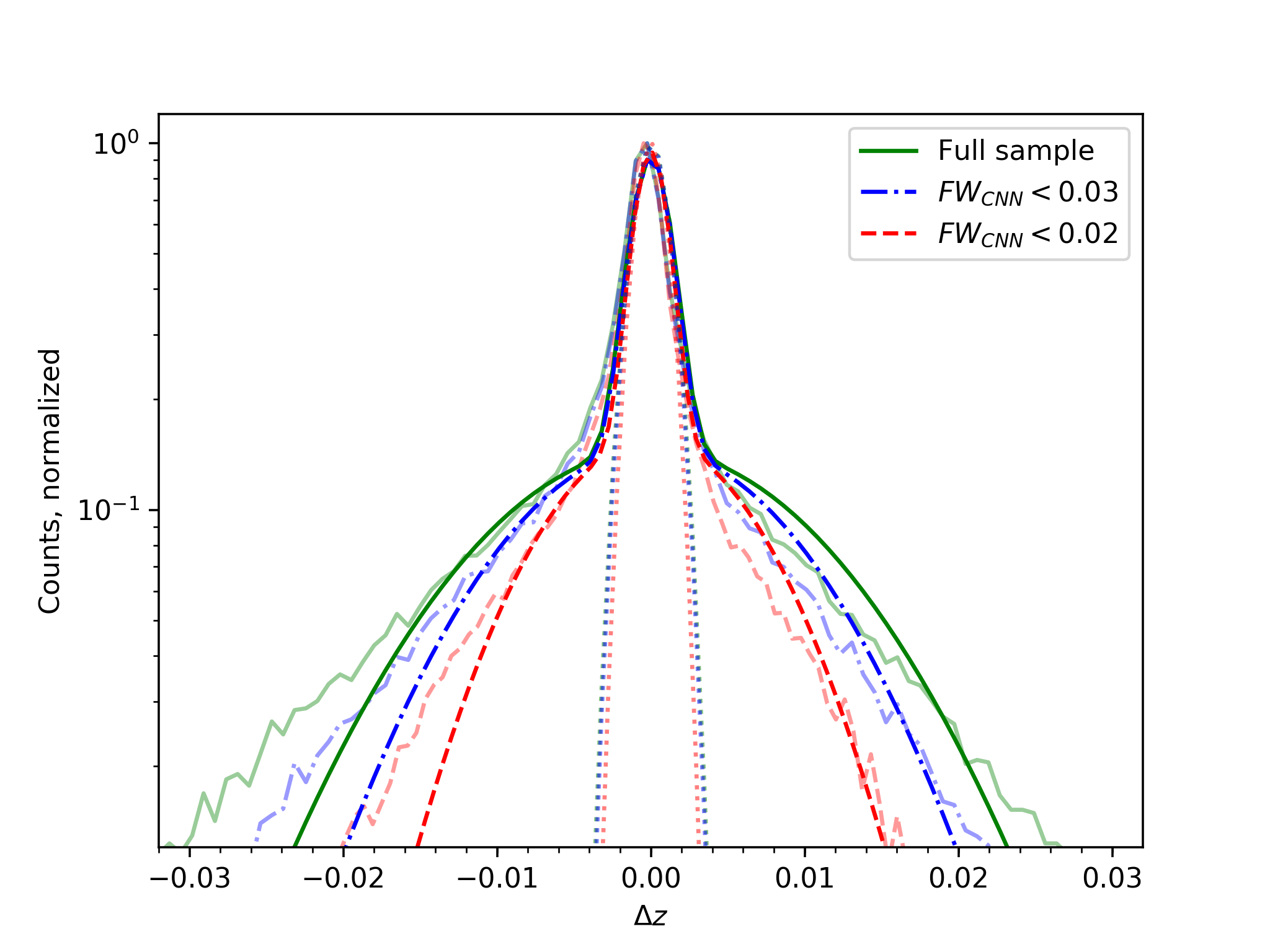}
\caption{\zw uncertainties (thin lines) modeled with a double-Gaussian fit (thick lines) for the full sample (green) and for galaxies with CNN PDF width$<$0.03 and 0.02 (blue and red, respectively). The dotted lines represent the  $\sigma_S$ of the small-scale component fit.
              }
\label{two-gauss-fit}
\end{figure}

The distribution of the \zw uncertainties are clearly non-Gaussian,  with a high compact core and a lower and broader component, as shown in Figure~\ref{two-gauss-fit}. 
\citet{AragonCalvo2015} proposed  modeling the \zw errors with a double-Gaussian function that reflects  the small-scale and large-scale errors,
\begin{equation} \label{2gauss}
f(\Delta z) = C_{\text{S}} \exp \left[ -\dfrac{(\Delta z)^2}{2\sigma^2_{\text{S}}} \right] + C_{\text{L}} \exp \left[ -\dfrac{(\Delta z)^2}{2\sigma^2_{\text{L}}} \right],
\end{equation}
where $C_{\text{S}}$, $C_{\text{L}}$, $\sigma_{\text{S}}$, and $\sigma_{\text{L}}$ are the normalization coefficients and standard deviations for the small-scale and large-scale component, respectively.
The result of the fit of Eq.~\ref{2gauss} to the uncertainties distribution of \zw based on the median is presented in Figure~\ref{two-gauss-fit} for the full galaxy sample and for galaxies with CNN PDF width $<$0.03 and 0.02. 
The fitted values for the full sample are $C_{\text{S}} = 1.0, \hspace{2mm} \sigma_{\text{S}} = 0.00122, \hspace{2mm} C_{\text{L}} = 0.1$, and $\hspace{2mm} \sigma_{\text{L}} = 0.01038$.
%
The small-scale redshift error dispersion $\sigma_{\text{S}}$ corresponds to a distance uncertainty of $\sim 5$ Mpc, of the order of the CW reconstruction uncertainties and non-linear processes (e.g.,  peculiar velocities), while the large-scale error dispersion $\sigma_{\text{L}}$ corresponds to $\sim 46$ Mpc, similar to the \zcnn uncertainty. 
Selecting galaxies with smaller CNN PDF widths leads to an improvement of the  large-scale \zw errors, while the small-scale component is almost unchanged. The values of $\sigma_{\text{S}}$ are reported in Table~\ref{tab:trials}.  

\begin{figure*}
\centering 
   \includegraphics[width=\hsize]{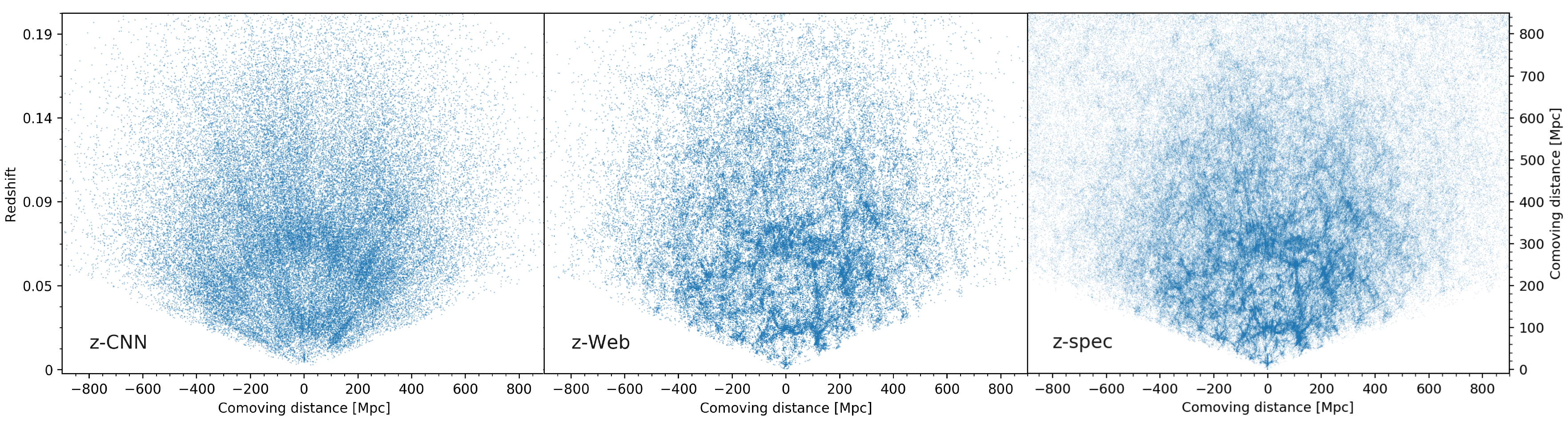}
   \caption{Galaxy distribution based on \zcnn (left panel), \zw (median, central panel), and spectroscopic redshifts (right panel) with z$\le$0.2. The 2D projections include galaxies with $0^{\circ} <\delta< 45^{\circ}$ and $109^{\circ}<\alpha<264^{\circ}$.
   }
   \label{fig:cw_zweb-zcnn}
\end{figure*}
 Finally, the galaxy distribution obtained with the \zw (median) and the \zcnn redshifts are compared in Fig.~\ref{fig:cw_zweb-zcnn}. The \zweb method performs as expected: the prior information of the spectroscopic CW density field places more galaxies inside the structures, significantly enhancing the CW features, which are barely seen with the CNN redshifts. This clearly illustrates the benefit of adding spatial information from spectroscopic surveys.
\subsection{Statistical behavior of the \zweb PDFs and redshift point estimates}  
\begin{figure}
\centering
\includegraphics[width=\hsize]{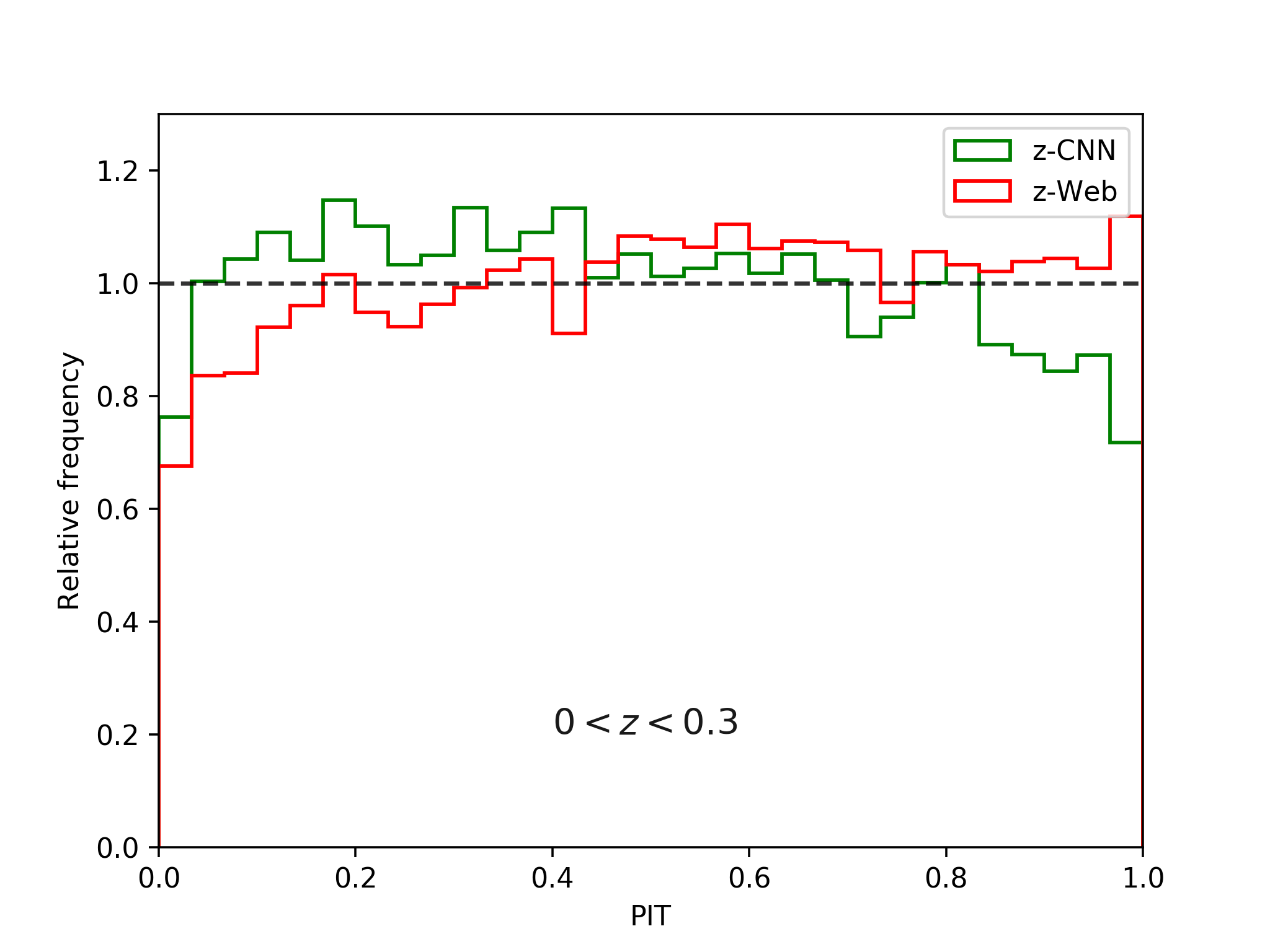}
\caption{Probability integral transform distribution of the CNN PDFs (green histogram) and \zweb PDFs (red histogram) in $0< z < 0.3$.
        }
\label{fig:PIT}
\end{figure}

 The final PDF is significantly modified with respect to the original CNN PDF. 
 We assess the predictive performance of the \zweb PDFs using the PIT test (see Section \ref{subsec:photozMGS}). The PIT distributions are presented in Fig.~\ref{fig:PIT} in the redshift interval $0<z<0.3$, where the majority of our galaxies reside. The PIT distribution is shown for the CNN PDFs (green) and the \zweb PDFs (red). As do the CNN PDFs, the \zweb PDFs exhibit a nearly flat distribution indicating that they are also well-calibrated 
 probability distributions, providing a reliable estimate of the redshift uncertainty. 
 However, they are not exempt from a small bias since a slope is observed, which indicates a slight underestimation of the \zweb redshifts.
 
 
\begin{figure}
\centering
\includegraphics[width=\hsize]{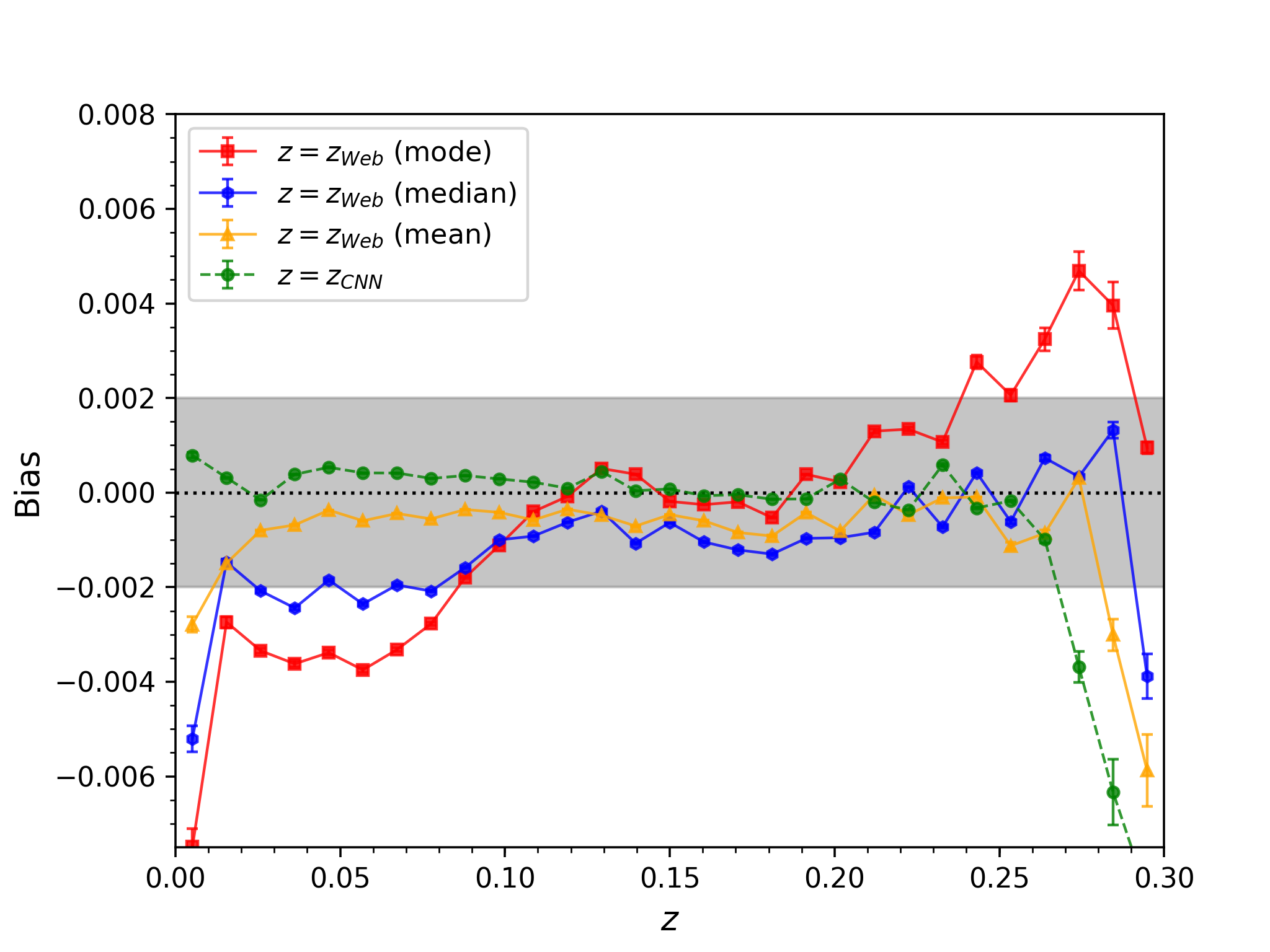}
\caption{Mean of the residuals (or bias, $<\Delta z>$) as a function of photometric redshift for the \zw (mode: red, median: blue, mean: yellow) and \zcnn (green). The gray-shaded region ($<\Delta z> < 0.002$) is the maximum bias requirement for the Euclid mission in all the photometric redshift bins.
        }
\label{fig:bias}
\end{figure}

 Future cosmological missions request strong constraints on the maximum redshift bias (defined as the mean residual; see Section \ref{subsec:photozMGS}) in photometric redshift bins used for the tomographic analyses. In particular, 
 the bias requirement for the Euclid mission is $\langle\Delta z\rangle<$0.002 \citep{Knox2006}.
 Figure~\ref{fig:bias} shows the bias, $\langle{\Delta z}\rangle$, as a function of \zw defined as the mode, median, and mean of the PDF and \zcnn, while the gray-shaded region shows the maximum bias requirement.
  The mean redshift estimates based on the CNN and \zweb PDFs show a very small bias at all redshifts, fully within the expected cosmological constraint.
 The \zw median redshift shows a small bias still within the constraint, but it appears marginally consistent at low z. However, when using the mode the bias exceeds the tolerance region below $z = 0.10$ and above $z = 0.25$. The anchor onto the main peak of the density field 
 makes this \zw estimate less robust for cosmological use. 

\begin{figure}
\centering
\includegraphics[width=\hsize]{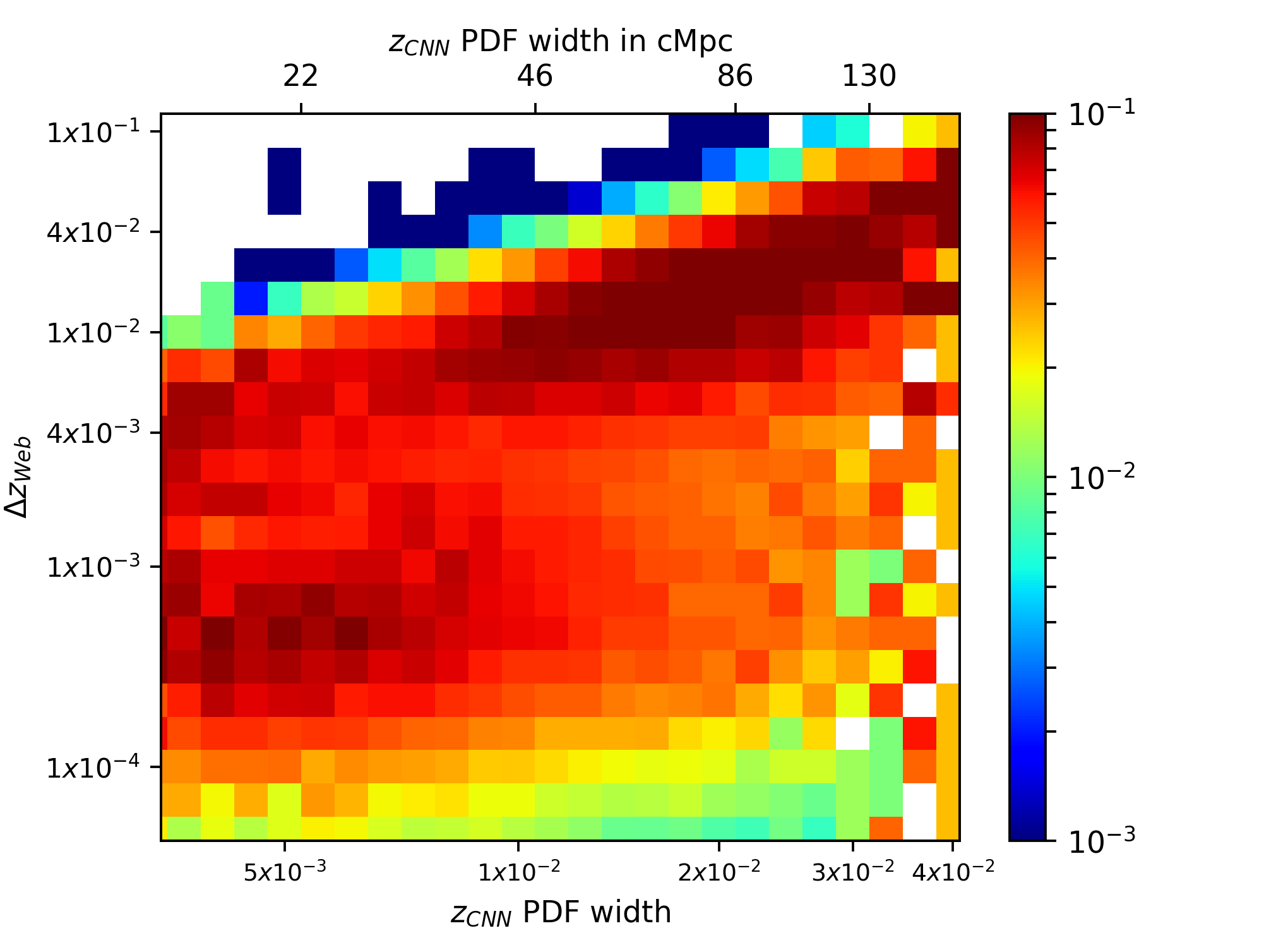}
\caption{Two-dimensional distributions of the \zw residuals (mode) as a function of CNN PDF width. The histograms are normalized by the area separately for each bin of PDF width. The narrower CNN PDFs properly enclose the true redshift, allowing the boost in redshift accuracy for most sources when combined with the CW information.  
       }
\label{fig:log_err}
\end{figure}
%
\subsection{Impact of the initial CNN PDF width}
 The performance of the \zweb method depends on the quality of the initial CNN PDF. If the PDF is narrow enough so it encompasses only a few CW structures, then it increases the probability of finding the structure the galaxy belongs to. 
 In Fig.~\ref{fig:log_err} we show the \zw residual (mode) as a function of the CNN PDF width. 
 First, the global trend is that the accuracy of \zw improves when the PDF width gets narrower, which is also the case of the \zcnn (Table~\ref{tab:trials}). This reflects the reliability and unbiased behavior of the CNN PDF.
  Then, when the CNN PDF width is narrower than a characteristic scale, or $\sim$80 cMpc, the fraction of greatly improved \zw ($\Delta z\le 0.002$) increases and becomes the majority for galaxies with a width $\le0.01$, at which point no further mismatches between structures are possible.   
  The improvement induced by the \zweb method when restricting the sample to CNN PDF widths lower than $\sigma_{\text{CNN}}$=0.03 and 0.02 are reported in Table~\ref{tab:trials}. All the statistical numbers improve, 
  in particular the $\sigma_\text{MAD}$ decreases by almost a factor of 2 and the fraction of galaxies with residuals lower than 10cMpc is higher than 55\%.  
  
  
%
\subsection{\zweb performance with respect to galaxy properties}
 In this section we examine the performance of the \zw for different categories of objects, such as star-forming versus passive galaxies, or as a function of group membership since the prior knowledge from the spectroscopic density field may impact different populations differently.
\subsubsection{Galaxy type}
\begin{figure}
\includegraphics[width=\hsize]{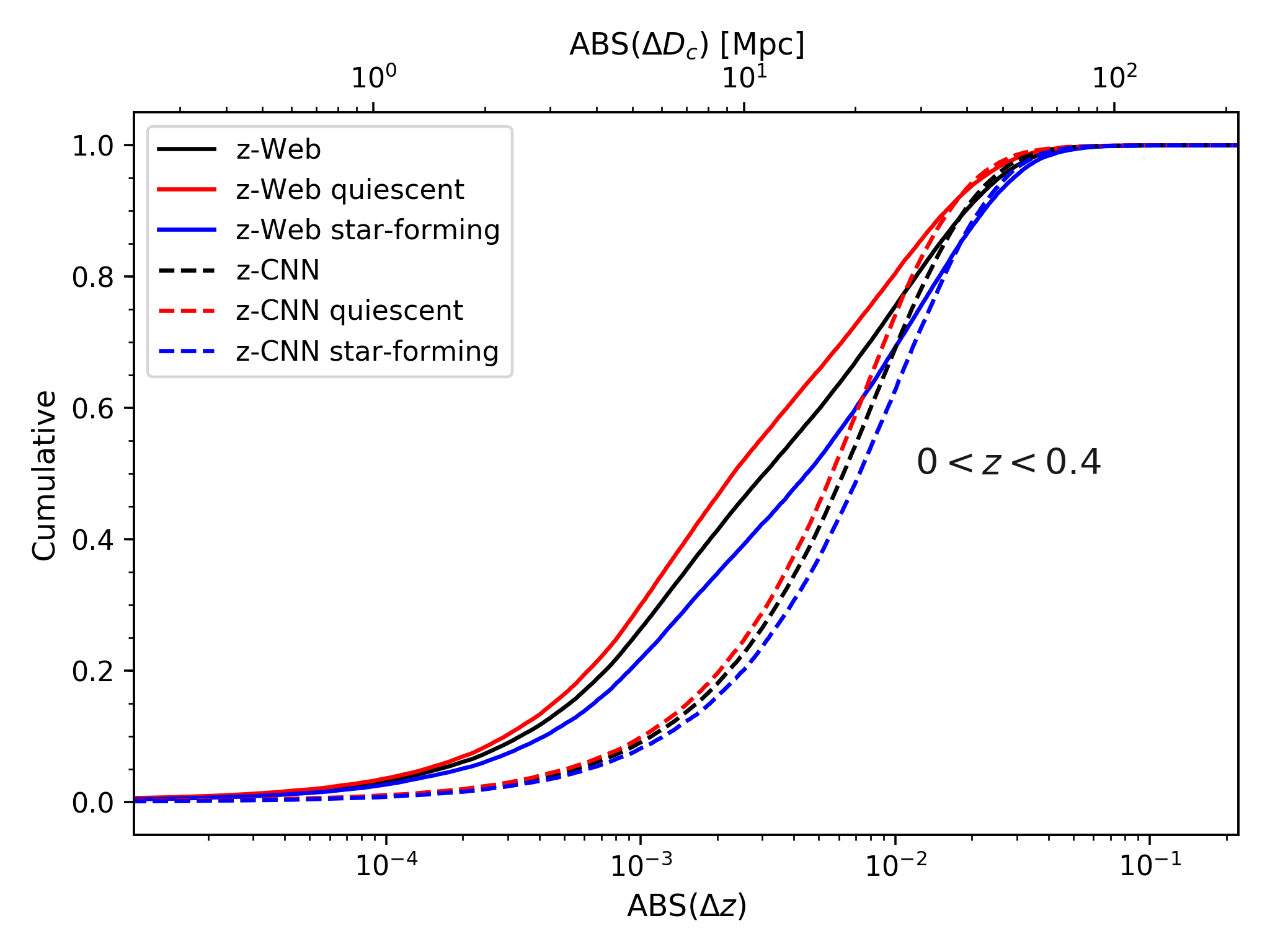}
\caption{Cumulative distribution of the residuals for quiescent galaxies (red), star-forming galaxies (blue), and the whole population (black) with \zcnn (dashed lines) and \zw (median, solid line). }
\label{hist-cdf-sfr}
\end{figure}
 We split the active and passive MGS galaxies according to the specific star formation rate values measured by \citet[]{Brinchmann2004}. We consider active galaxies as those with $\log({\rm sSFR})\ge -11$ and passive otherwise. Figure~\ref{hist-cdf-sfr} shows the cumulative distributions of the \zw and \zcnn residuals for the two subsamples.
 Passive galaxies show a better redshift accuracy than active ones with the CNN method. This could 
be due to the brighter magnitude distribution of passive galaxies, but also  to the greater diversity of star-forming galaxies (e.g., due to clumpiness, dust lanes, inclination), making the deep learning technique slightly less efficient.
After applying the \zweb method, passive galaxies have a greater boost in accuracy than active galaxies. This is a natural consequence of the biased distribution: passive galaxies 
 are preferentially in the high-density regions of the CW compared to star-forming galaxies. This segregation effect was recently quantified with respect to filaments in spectroscopic \citep[]{Malavasi2017, Kraljic2018} and photometric \citep[]{Laigle2018} surveys. At least at low redshifts, passive galaxies are statistically closer to filaments than active ones at similar stellar mass. The \zweb method is therefore expected to be more effective for the former population.  
\subsubsection{Group membership}
\begin{figure}
\centering
\includegraphics[width=\hsize]{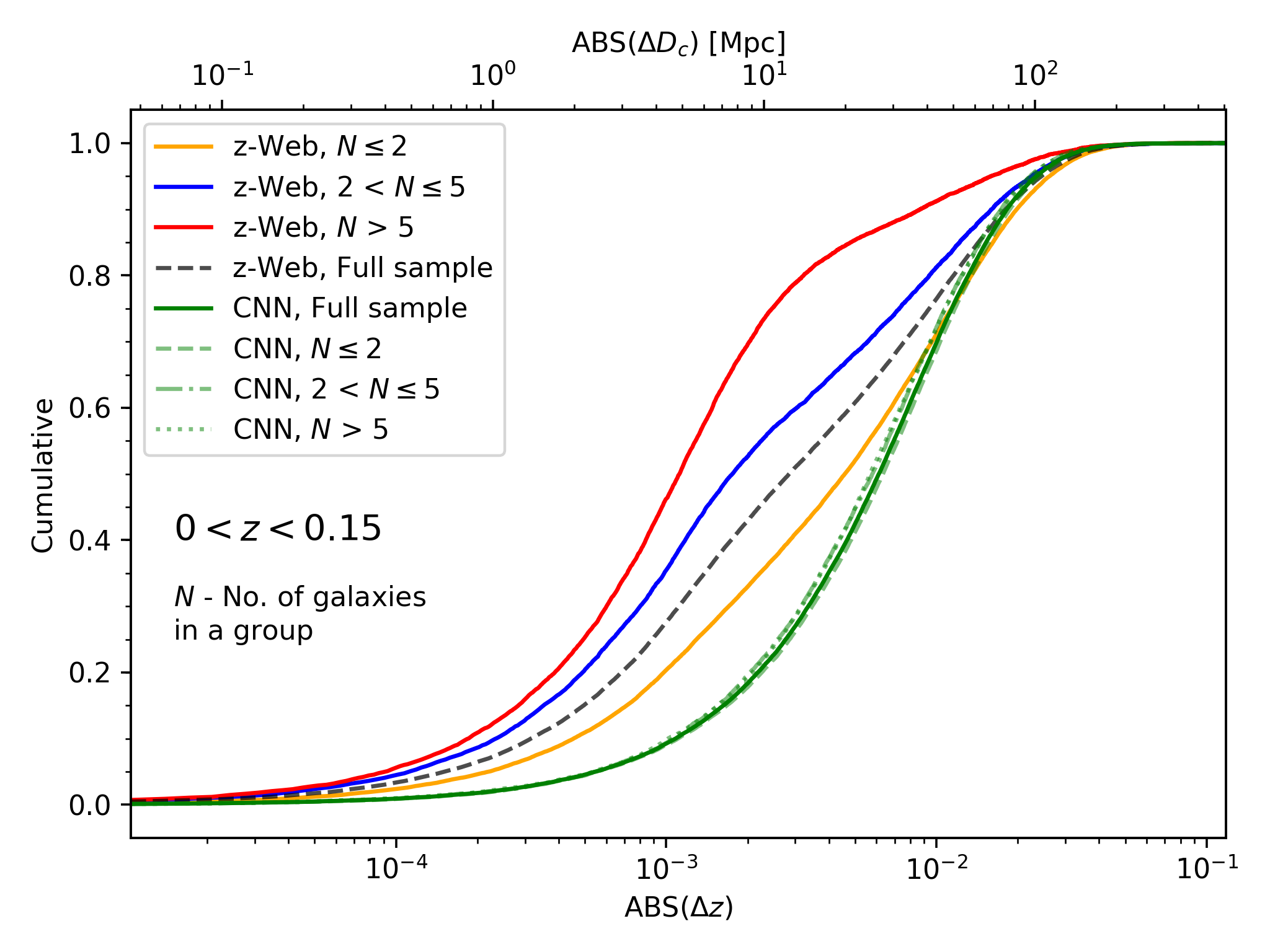}
\caption{Cumulative distribution of the residuals \zw (colored lines) and \zcnn (green lines) for galaxies belonging to different group sizes: one or two members (orange), three to five members (blue), and higher (red).}
\label{cdf-groups}
\end{figure}
%
Almost half of the galaxies in the local universe are part of gravitationally bound systems. These groups are distributed along or at the intersection of the filaments of the CW, and represent the peaks of the galaxy density field. These peaks,  however, are  slightly diluted along the line of sight due to the peculiar velocities of the galaxies that introduce redshift-space distortions (Fingers of God), which are not corrected for before reconstructing the local density field in the present work. Since most of the groups have a velocity dispersion lower then $\sigma_v=600$  km s$^{-1}$ \citep[]{Tempel2014}, it will impact the radial distance by less than 10 cMpc, i.e., 4 times lower than the current accuracy achieved by the \zcnn redshifts.  
We match our sample of MGS galaxies with the group catalog of \citet[]{Yang2007}, based on a friends-of-friends algorithm performed up to $z=0.2$. The MGS galaxy sample is then split according to group size and the redshift residuals for the different subsamples are shown in Fig.~\ref{cdf-groups}. As expected, the performance of the \zweb method is highly dependent on the number of group members: the boost in accuracy is most prominent for galaxies belonging to large groups. For the largest group sample ($N>5$), $\sim 98\%$ have improved \zw compared to \zcnn, and $80\%$ have an error smaller than 10 Mpc. The improvement remains significant for galaxies in groups of intermediate size ($3<N\le 5$), with $\sim 90\%$ having better \zw than \zcnn, compared to $\sim60\%$ for isolated or in paired galaxies. On the contrary, the \zcnn residuals are identical for all the subsamples since no prior knowledge about group membership is specified.
\subsection{Impact of the spectroscopic CW reconstruction}
\begin{figure}
\includegraphics[width=\hsize]{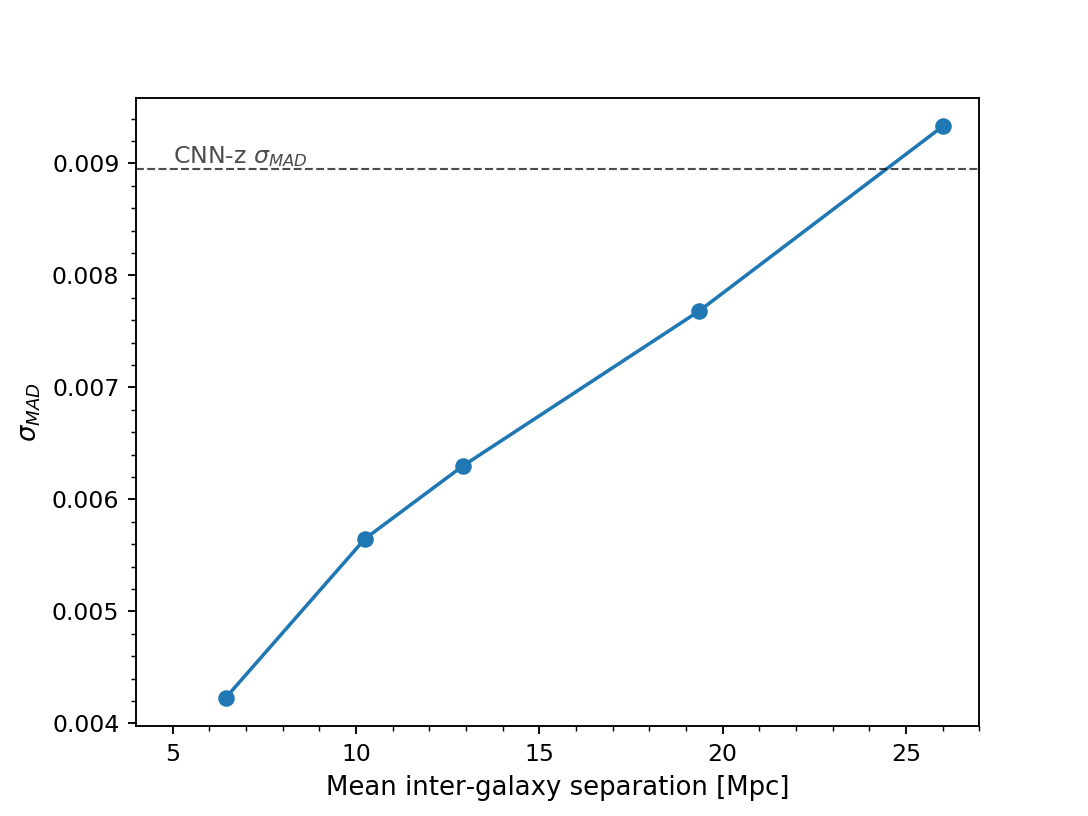}
\caption{Mean absolute deviation, $\sigma_\text{MAD}$, of the \zw (based on the median) as a function of the mean intergalactic separation of several sparse spectroscopic samples. The black horizontal line shows the $\sigma_{\text{MAD}}$ of the original \zcnn.
              }
\label{fig:mad_dinter}
\end{figure}

 The spectroscopic sampling of the galaxy density field is the second most important  ingredient of the \zweb method after the quality of the original PDF. We evaluate its impact by randomly reducing the number of galaxies in the spectroscopic survey by several factors (25, 12.5, 3.7, 1.6\%). 
 We restrict this analysis to $z\le 0.15$, where the mean intergalactic separation varies slowly with z (Fig.~\ref{fig:sample}). The \zweb method is 
  reapplied using the density fields and CW features computed for each of the sparse spectroscopic samples. In Fig.~\ref{fig:mad_dinter}, the rms of the median \zw residuals, $\sigma_{\rm MAD}$, are shown for the different subsamples, corresponding to different mean intergalactic distances. Decreasing the sampling decreases the performance of the method, but it takes a very sparse sampling to reach the rms value of the original \zcnn. 
 It can also get worse when the poor reconstruction of the galaxy density field  systematically misidentifies the structures that galaxies belong to.
%
%
\subsection{\zweb performance for the GAMA survey}
\begin{figure*}
\centering 
   \includegraphics[width=0.45\hsize]{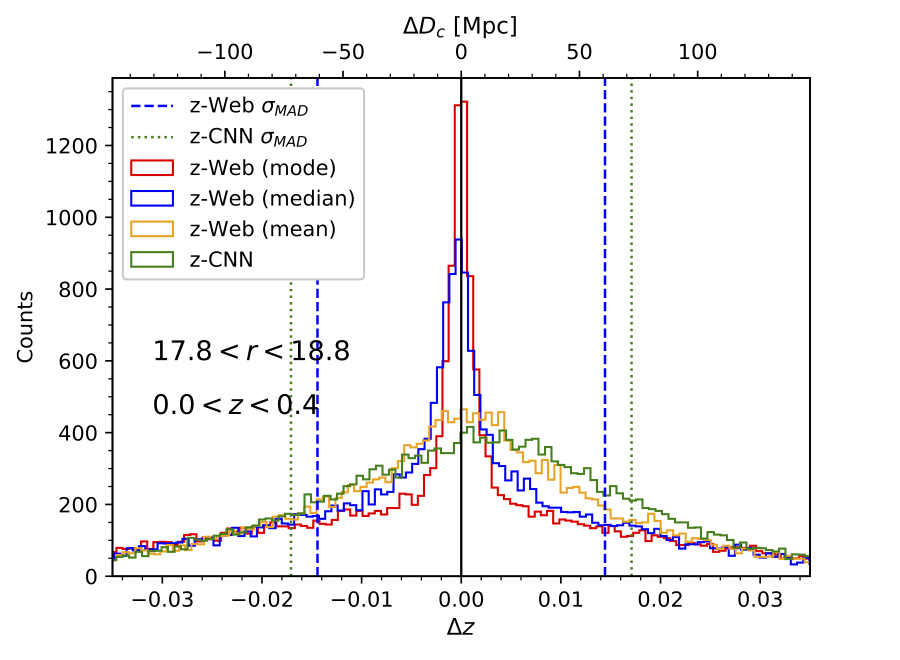}
   \includegraphics[width=0.45\hsize]{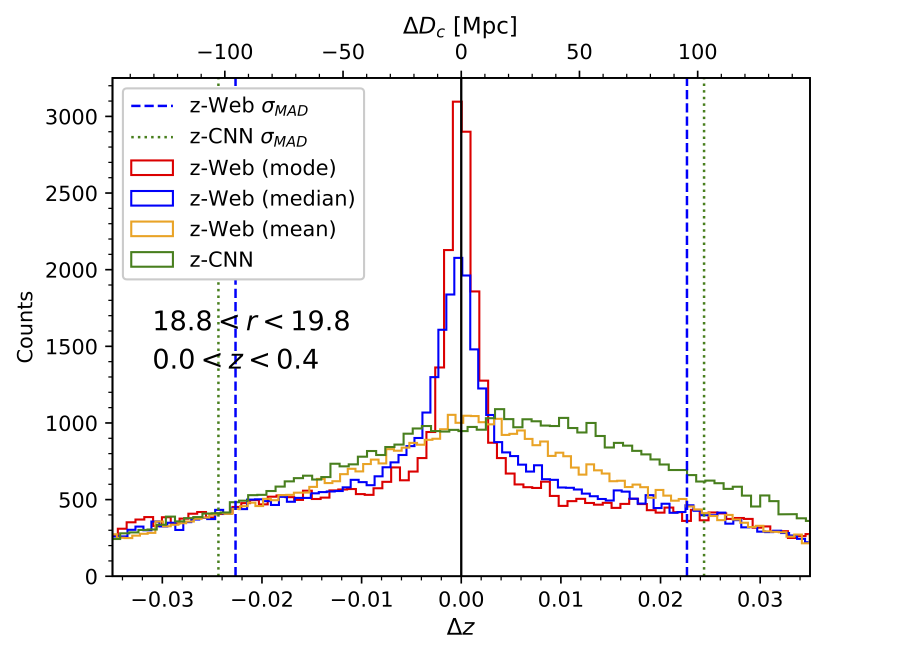}   \includegraphics[width=0.42\hsize]{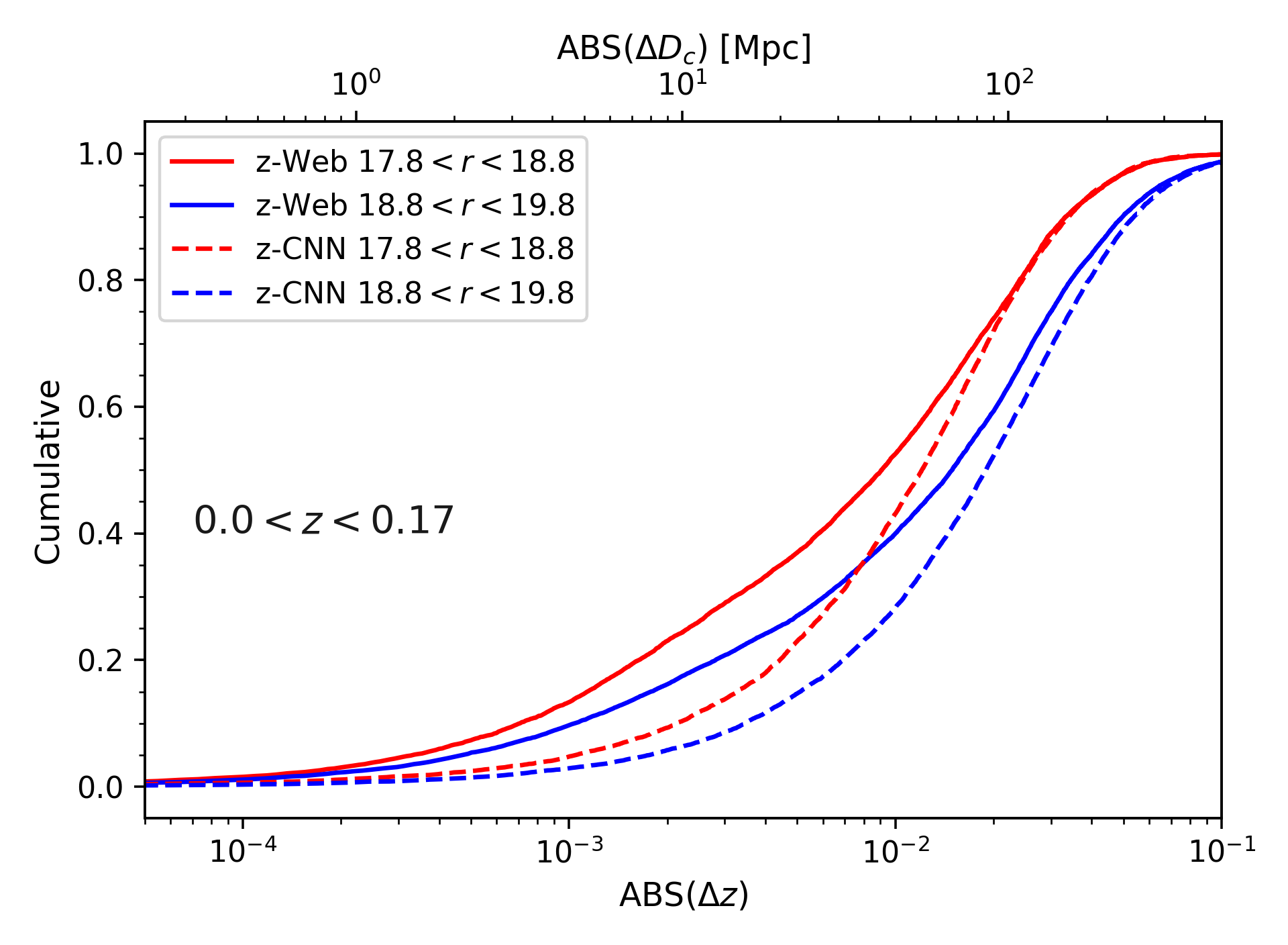}
   \includegraphics[width=0.42\hsize]{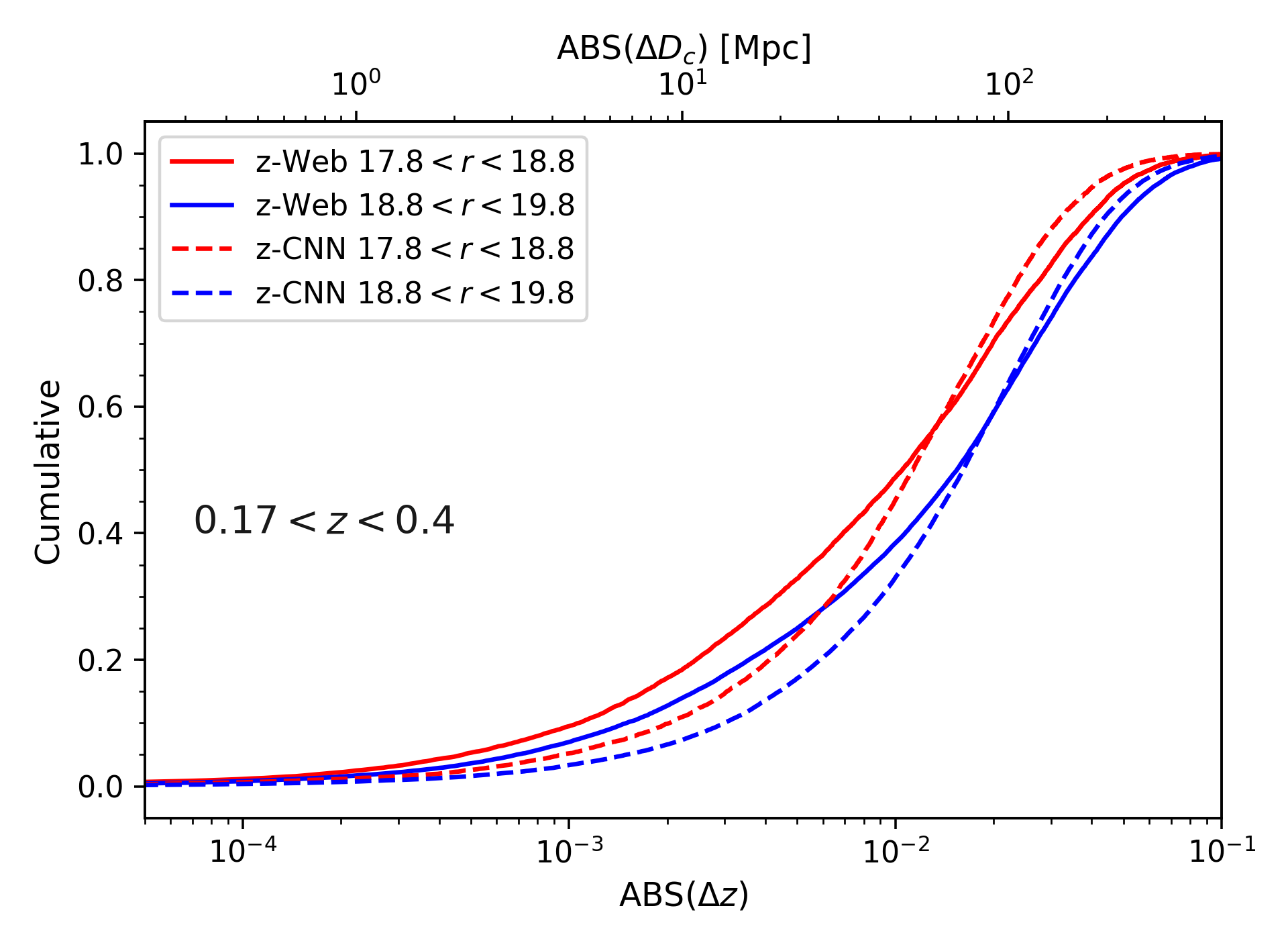}
\caption{Redshift residuals of the \zcnn and \zw for the GAMA survey. Top: Differential histograms with the full sample for the \zcnn (green) and the \zw (with the mode, median, and mean of the PDF) at $17.8\le r\le 18.8$ (left), and $18.8\le r\le 19.8$ (right). 
Bottom left: Cumulative histograms for the \zcnn (dashed line) and \zw (median only, solid line) at low redshift ($z\le 0.17$) and split into two magnitude bins (red: $17.8\le r\le 18.8$; blue:$18.8\le r\le 19.8$). Bottom right: Same as left,  but at high redshift ($0.17\le z\le 0.4$).
}
\label{fig:zweb-gama}
\end{figure*}

\begin{figure}
\centering
\includegraphics[width=\hsize]{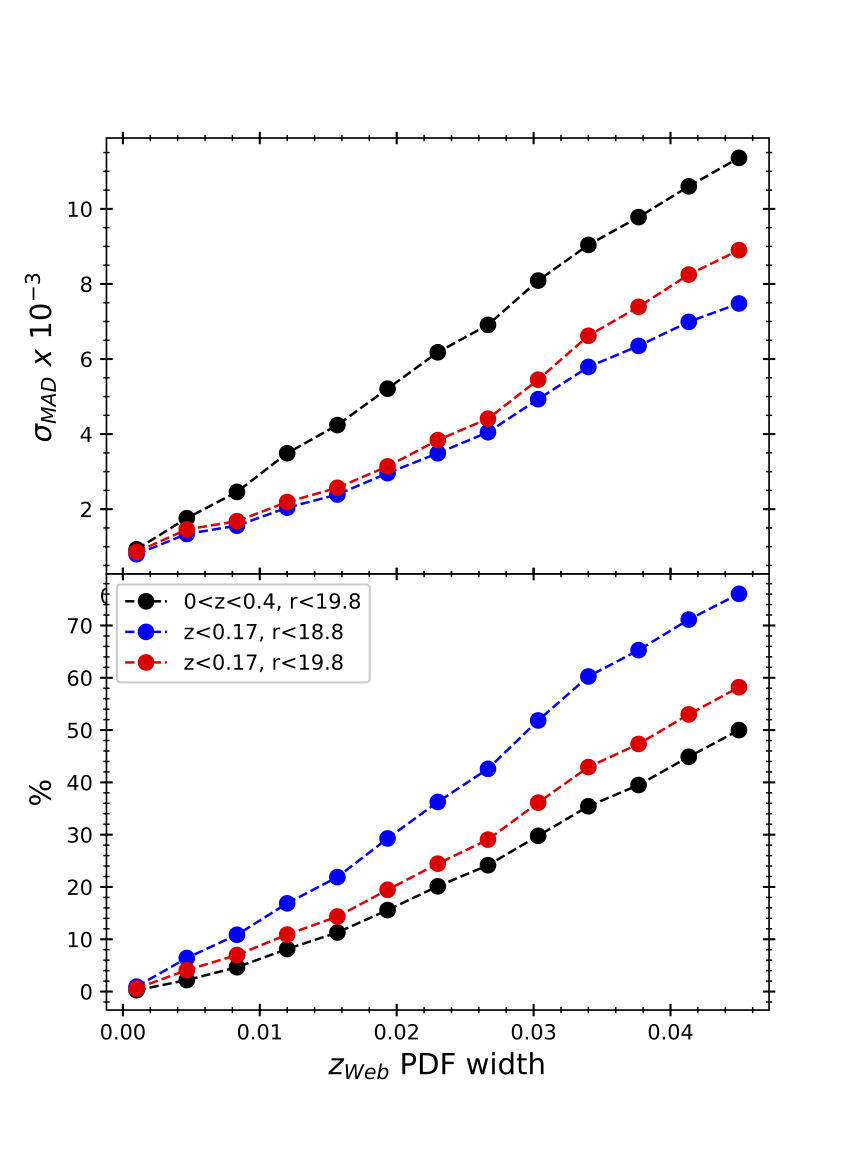}
\caption{ WEBz performance for GAMA galaxies selected according to their \zweb PDF width. Top panel shows the $\sigma_{\rm MAD}$ for the whole population ($0<z<0.4$ and $r<19.8$; black), the low-z bright ($z<0.17$ and $r<18.8$; blue), and the low-z faint ($z<0.17$ and $r<19.8$; red) subsamples. The bottom panel shows the cumulative fraction of galaxies for each sample. 
      }
\label{fig:GAMA_MAD-vs-PDFWidth}
\end{figure}

In this section we push the \zweb method to the fainter galaxy population of the GAMA survey. The CW from the SDSS-BOSS spectroscopic survey is combined with the \zcnn photometric redshifts of GAMA. As mentioned in Section 2, the low S/N of the SDSS images for GAMA sources ($17.8\le r \le19.8$) leads to wider \zcnn PDFs than for the MGS sample. The impact on the performance are summarized in Fig.~\ref{fig:zweb-gama} and Table~\ref{tab:gama}. For the whole sample, the \zw residuals still show a high fraction ($\sim$70\%) of improved photometric redshifts well centered at $\Delta z$=0, while the \zcnn residual appears slightly biased ($<\Delta z>\sim 0.005$; Fig~\ref{fig:zweb-gama}, left panel). 
 In Fig~\ref{fig:zweb-gama}, bottom left and right, we distinguish between low and high redshifts to partly disentangle the impact of the CW reconstruction from the \zcnn PDF widths. At low $z$ (bottom left panel), where the CW is better reconstructed, the \zw are better than the \zcnn in  both the bright and faint magnitude bins. It more than doubles the number of galaxies with distance uncertainties better than 10cMpc (see Table~\ref{tab:gama}).
 At higher redshift (bottom right panel), about half of the galaxies have improved photometric redshifts in both magnitude bins with a modest gain of highly accurate redshifts ($\le$10Mpc). 
 The degradation for the second half of \zw with respect to \zcnn can be attributed to the sparse CW reconstruction, which introduces associations with the wrong density peaks (as mentioned in section 4.5). \\
 As shown in Table~\ref{tab:gama}, when restricting the sample to the narrowest \zcnn PDF widths, the $\sigma_{\rm MAD}$ for \zw still improves the accuracy 
 but only for a small fraction of the objects.  More practically, we can select a population with a desired redshift accuracy based on their final \zweb PDFs, despite their more complex shapes. In Fig~\ref{fig:GAMA_MAD-vs-PDFWidth}, we show the evolution of the \zw accuracy as a function of the \zweb PDF width and the relative fraction of galaxies considered. The accuracy deteriorates progressively toward higher PDF widths.   
  With a cut at PDF width $\le$0.04 for the low-z sample ($z \le 0.17$), we can select 70\% (50\%) of galaxies brighter than 18.8 (19.8), with an accuracy better than $\sigma=0.007$ (0.008).  
 For the whole GAMA sample, you can select a population with $\sigma=0.01$ by applying a PDF width cut of 0.038, which will enclose 40\% of the population.
 In conclusion, the method still benefits the photometric redshifts of galaxies two magnitudes fainter than the spectroscopic sample used to reconstruct the density field. 
%
%
%
\begin{table}
\caption{Performance of \zweb (median) in GAMA survey for $17.8 < r < 18.8$ (top) and $18.8 < r < 19.8$ (bottom), low-redshift samples and different CNN PDF width selections.Values are reported for \zw/\zcnn. } 
\label{tab:gama}      
\centering          
\begin{tabular}{|  c | c | c | c | c |}  \hline       
GAMA         &$\sigma_{\text{MAD}}$  &$\eta$& \multicolumn{2}{|c|}{$\Delta$\zw$<$} \\
number of    &  ($\sigma_S$)    &      &      10cMpc    &    $\Delta$\zcnn\\
galaxies     &  $\times 10^{-3}$&  \%  &   \%           & \%    \\\hline \hline
\multicolumn{5}{|c|}{ $ 17.8 < r < 18.8$ } \\ \hline 
Full sample &  14.4/17.1 &  4.0/2.8 &  27/15  & 69 \\ 
   23,987   & (1.5)& &  &    \\  \hline
$z<0.17$ &  13.4/16.8 &  3.2/3.1 &  25/10  & 80 \\ 
   11,457   & (1.5)& &  &    \\  \hline
Width$<$0.03&  6.5/9.5  & 0.5/0.1  &  35/19  & 81 \\  
   3,513    & (1.6) & &   &    \\ \hline
Width$<$0.02&  4.7/6.5  & 0/0     &  41/26  & 92 \\  
    275     & (1.1) &    &  &    \\ 
\hline\hline
\multicolumn{5}{|c|}{ $ 18.8 < r < 19.8$} \\ \hline 
Full sample & 22.6/24.4  & 9.8/8.2  & 18/9   & 72 \\ 
 65,277     &(1.9) &  &   &    \\ \hline
$z<0.17$ &  21.1/25.4  & 9.9/12.0 &  18/7  & 97 \\ 
   16,872   & (1.6)& &  &    \\  \hline
Width$<$0.03& 7.62/10.7  & 1.6/1.6  & 28/18   & 69 \\
        242 &(1.6) & &  &    \\
\hline \hline
\end{tabular}
\end{table}
%

%

\section{Conclusion}
In this work, we revisited and extended the  study  of \citet[]{AragonCalvo2015}, who illustrated the benefit of combining photometric redshift PDFs with the knowledge of the CW to boost their accuracy. 

Here we make use of the robustness of the cosmic web extractor  DisPerSE, and the   more accurate and better calibrated photometric redshifts PDFs based on  a CNN. The density field and the main components of the cosmic web (nodes, filaments, and walls) are reconstructed with  DisPerSE,  up to 
$z\sim 0.4$  using the combined SDSS-MGS and BOSS surveys. The final PDF of each galaxy is obtained by combining  their original CNN PDF with the density field  and the distance to any CW structures along their line of sight, providing a new estimate of the photometric redshift, \zw. 
 We first apply this technique to galaxies from the MGS sample ($r\le 17.8$). Our main conclusions are as follows:
%
\begin{itemize}
\item For the whole population, the method improves the precision of $\sigma_{\rm MAD}$ by a factor of up to 2.5. By using the mode of the final PDF as the new photometric redshift value, the initial distance uncertainty of $\sim$40 cMpc shrinks to $\sim$17 cMpc.
\item The \zw 
accuracy
is degraded for 10\% of the sources that are associated with the wrong structure. This effect  can be mitigated by using the mean of the \zw PDF rather than the mode, at the price of a lower \zw precision.
\item The nearly flat PIT distribution shows that the final \zw PDFs are also well calibrated  and reliable. Although a small bias is observed, it can be kept within the requirements of cosmological missions  by choosing the mean or median as redshift point estimates.  
 This allows us to select populations according to their \zw uncertainties.
\item As expected, the final \zw precision depends on the original \zcnn PDFs: the narrower the CNN PDF,  the lower the number of intercepted structures, the higher the boost in \zw accuracy. By selecting a \zcnn PDF width narrower than 0.02 (i.e., $\sim$90 cMpc), $\sigma_{\rm MAD}$ is reduced by a factor of $\sim$2 and the fraction of galaxies with residuals lower than 10 cMpc exceeds $\sim$50\%. 
\item The \zweb method performs better for passive galaxies, due to their higher luminosities (S/N) and stronger correlation with the densest regions of the density field compared to star-forming galaxies. 
\item Using an independent SDSS group catalog, we find that the distance error for 80\% of the galaxies in large groups ($N>5$) is smaller than 10 Mpc. 
\item Up to a mean intergalactic distance of 20 cMpc, achieved by reducing the sampling for the CW reconstruction, the \zweb method still improves the photometric redshifts.   
\end{itemize}
 We then extended the method to galaxies that are two magnitudes fainter than MGS ($r\le 19.8$) using the GAMA spectroscopic survey.
Despite the reduced size of the training sample and the lower S/N of the images, we were able to obtain a CNN photometric redshift precision of $\sigma_{\text{MAD}} < 0.026$. We apply the \zweb method in this faint regime while keeping the same CW information as above. We found the following:
\begin{itemize}
\item Although the CNN PDFs are significantly wider, 65\% of the \zweb redshifts are better than \zcnn and twice as many objects (i.e., $\sim$20\%) have distance uncertainties lower than 10 cMpc. However, the gain in $\sigma_{\rm MAD}$ is  only $\sim$10\%. Interestingly, the \zweb method allows us to get rid of a small bias observed for the \zcnn in the faintest magnitude bins.
\item While the \zcnn accuracy depends mainly on the S/N of the images rather than on redshift, a larger fraction of galaxies has improved \zw at low redshift ($z\le 0.17$) than at high redshift ($0.17<z<0.4$). This reflects the importance of the resolution of the CW reconstruction.
\item The \zweb PDFs can be used to select  galaxies with a desired redshift accuracy (e.g., galaxies with \zweb PDF width lower than 0.038 (40\%)  have an accuracy of $\sigma \sim 0.01$). This will be of interest when the \zcnn and \zw  are extended to $r\le 19.8$ for the entire SDSS, but with the drawback of a poorly controlled selection function. 
\end{itemize}

Combining the cosmic web with photometric redshift PDFs so as to anchor galaxies to the structures they are most likely to inhabit is a powerful method for improving the original photometric redshifts. The SDSS survey is particularly well suited for such an analysis as it combines highly accurate photometric redshifts ($\sigma \sim 0.01$) and good mapping of the cosmic web (with a resolution better that $\langle d_{inter} \rangle \le$10 cMpc). Attaching photometric galaxies to the  spectroscopically derived CW improves their photometric redshifts, even for galaxies one or two magnitudes fainter than the spectrocopic limit, as in the case of the GAMA survey.
 With this technique, constraining the environment of even faint galaxies is now within reach. This will enable extending galactic conformity inside groups \citep[]{Treyer2018}, for example,  or spin alignment \citep[]{Tempel2013} studies to the low-mass  galaxies.
     
The applications to future surveys are more tentative as the efficiency of the method depends first on the accuracy of the photometric redshifts and their associated PDF widths, and second on the resolution of the CW based on spectroscopic surveys on the same field.
 Multiband surveys like PAU \citep[with 30 narrowbands,][]{Eriksen19}, J-PAS  \citep[with 50 narrowbands,][]{Benitez2014}, 
 or low-resolution spectroscopy missions like SPHEREX \citep[]{Dore18} will deliver redshifts with uncertainties below $\sigma_z\le 0.01$, 
 but cosmological spectroscopic surveys like BOSS \citep[]{Dawson2016} and DESI \citep[]{DESI2016} have, or will have, a moderate resolution (with $\langle d_{inter}\rangle > 15-20$ cMpc), which will hamper the use of the \zweb method. 

On the other hand, CW mapping at high redshift with spatial resolution less than $\langle d_{inter} \rangle \sim 10$ cMpc is now within reach with VIPERS \citep[$z\sim 0.8$,][]{Guzzo14, Malavasi2017} or in the near future with PFS \citep[CW traced by the galaxies or by the gas with the tomography technique,][]{Tamura018}, as well as the spectroscopic survey modes of Euclid-Deep \citep[]{Laureijs11} and WFIRST \citep[]{Spergel2015} up to $z\sim 2.5$. 
However, at such a high redshift the current limitation  is the accuracy of the photometric redshifts.  Current techniques barely reach $\sigma\sim 0.03$ \citep[]{Moutard2016}, which again will restrict the use of the \zweb method. Extending the CNN method \citep[]{Pasquet2019} at higher redshift should allow us to pass this threshold. Preliminary CNN training on $ugriz$ CFHTLS images yields an accuracy below $\sigma\le 0.02$ at $i_{AB}\le 22.5$ (Treyer et al., in prep.) and $\sigma\le 0.015$ at $i_{AB}\le 23$ on deep HSC images combined with CLAUDS \citep[mimicking the LSST wavelength coverage and depth with ugrizY passbands;][]{Sawicki2019}. While very promising at intermediate redshifts ($z\le 1.5$), it is not yet optimal at higher redshift due to the poor spectroscopic training set and further improvements are needed to fully exploit the combination of the LSST survey with the Euclid and WFIRST missions. 
\begin{acknowledgements}
We thank the referee for useful comments.
This work has been carried out thanks to the support of the CNES, the  OCEVU Labex (ANR-11-LABX-0060), the Spin(e) project (ANR-13-BS05-0005, http://cosmicorigin.org), the DEEPDIP project (ANR-19-CE31-0023), the SAGACE project (ANR-14-CE33-0004) and the “Programme National Cosmologie et Galaxies” (PNCG).
\end{acknowledgements}

\section{Appendices} 
\begin{figure*}
   \begin{subfigure}[b]{0.5\textwidth}
            \includegraphics[width=0.83\hsize]{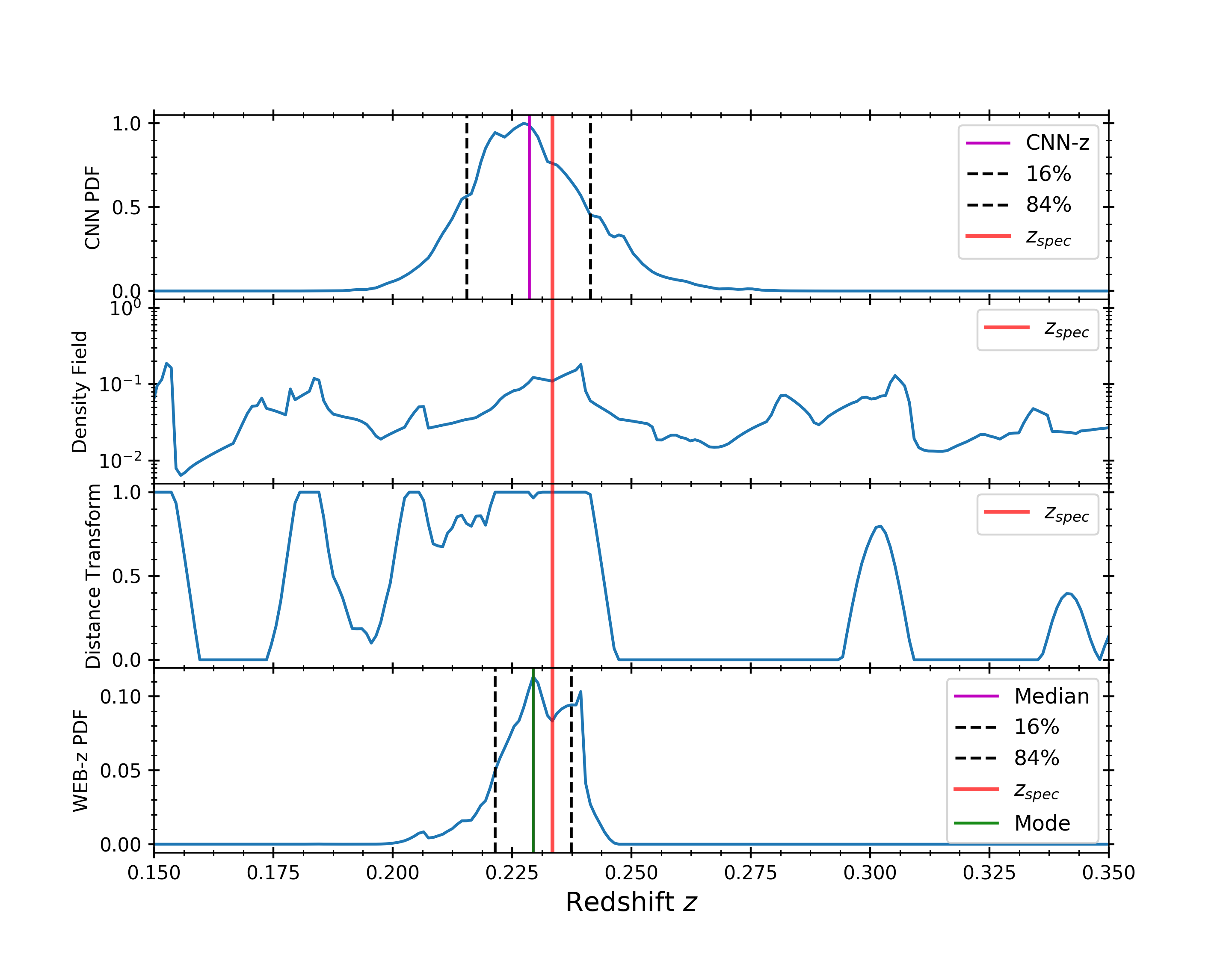}     
      \caption{ 
              }
         \label{pw-pdf-b}
   \end{subfigure}
   \begin{subfigure}[b]{0.5\textwidth}
            \includegraphics[width=0.83\hsize]{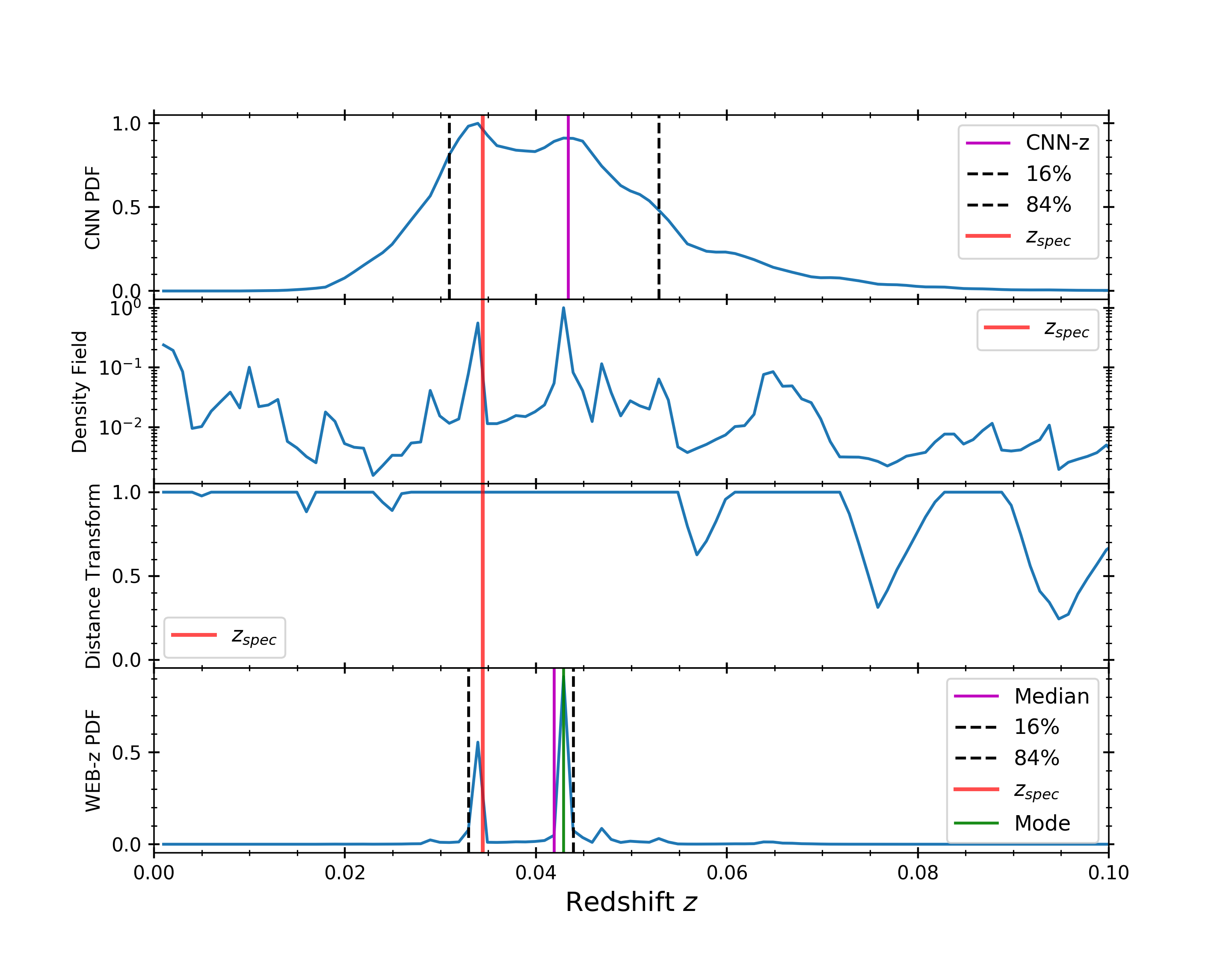}     
      \caption{ 
              }
         \label{pw-pdf-c}
   \end{subfigure}
   \begin{subfigure}[b]{0.5\textwidth}
            \includegraphics[width=0.83\hsize]{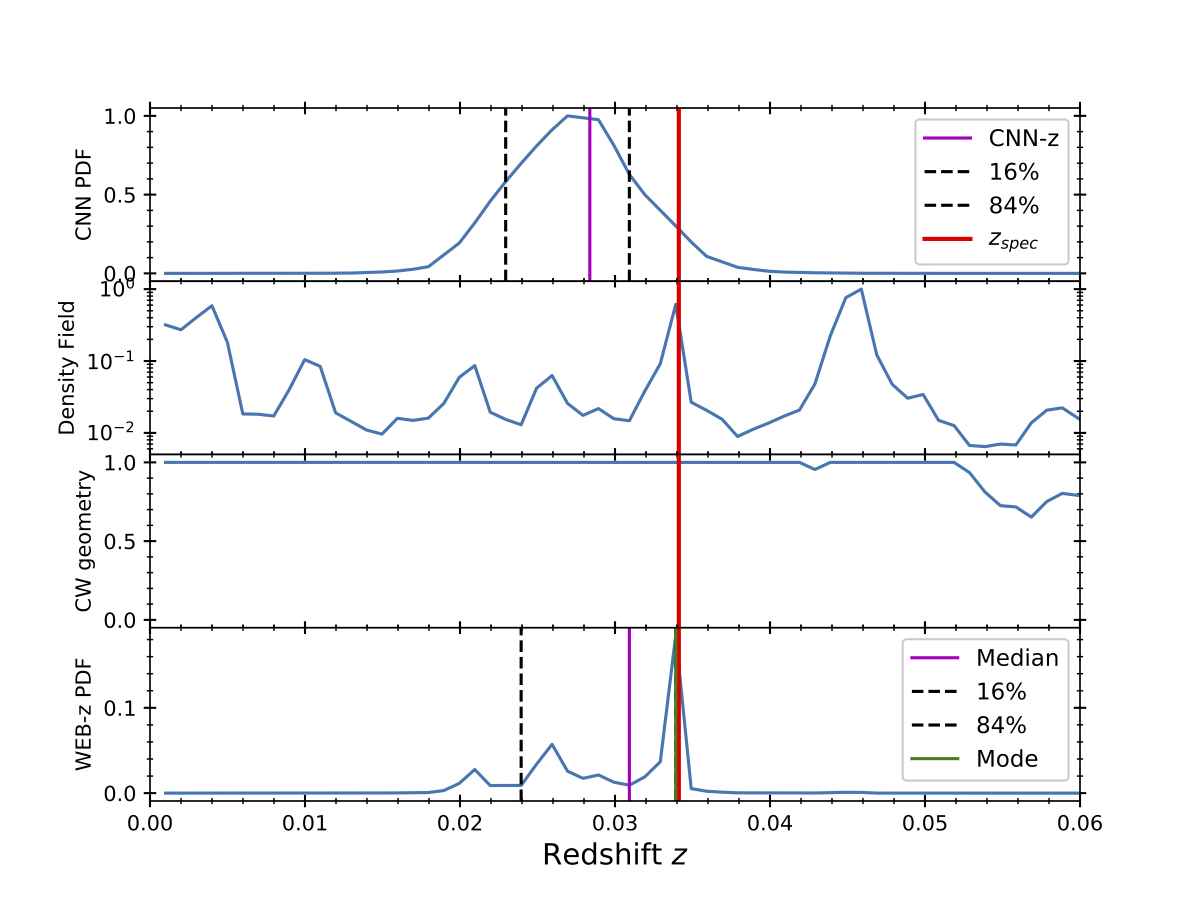}     
      \caption{ 
              }
         \label{pw-pdf-d}
   \end{subfigure}
   \begin{subfigure}[b]{0.5\textwidth}
            \includegraphics[width=0.85\hsize]{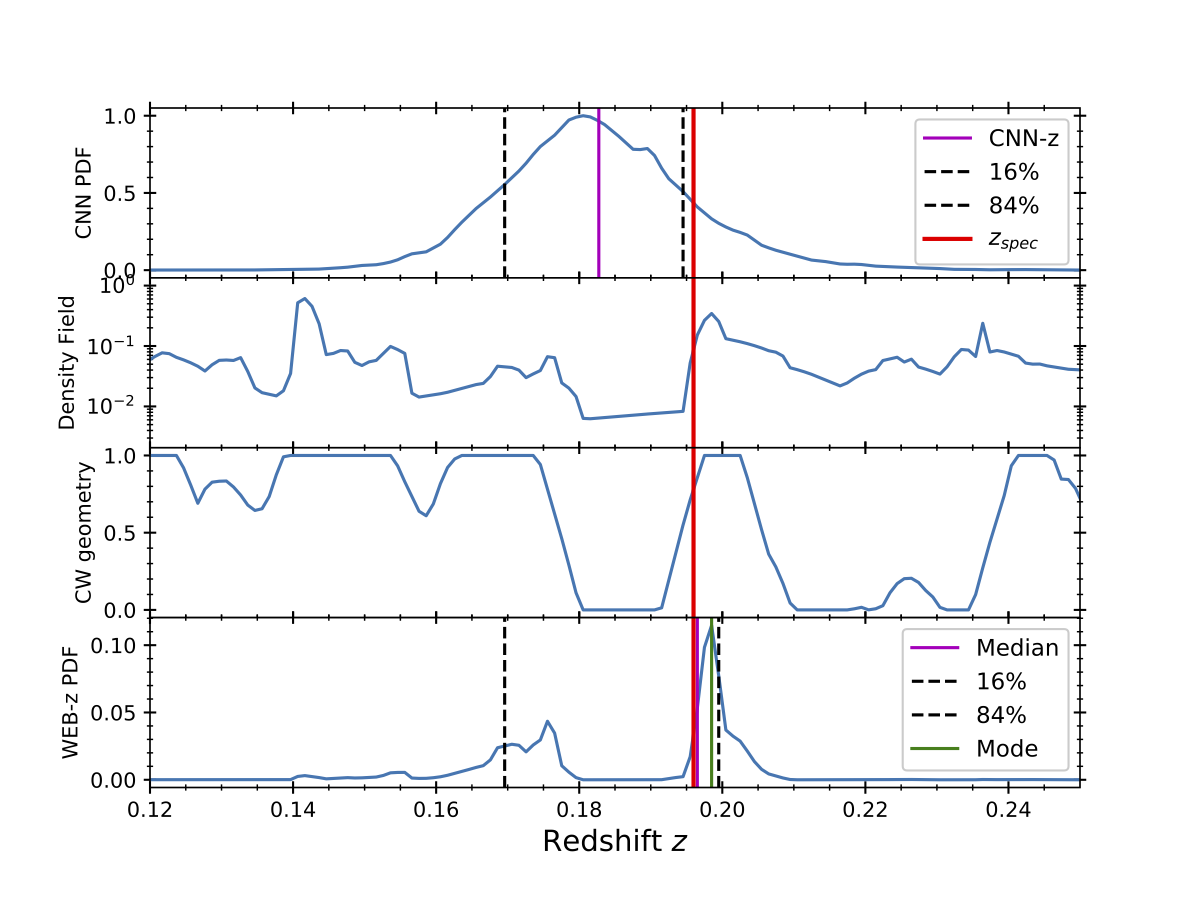}     
      \caption{ 
              }
         \label{pw-pdf-e}
   \end{subfigure}
   \begin{subfigure}[b]{0.5\textwidth}
            \includegraphics[width=0.85\hsize]{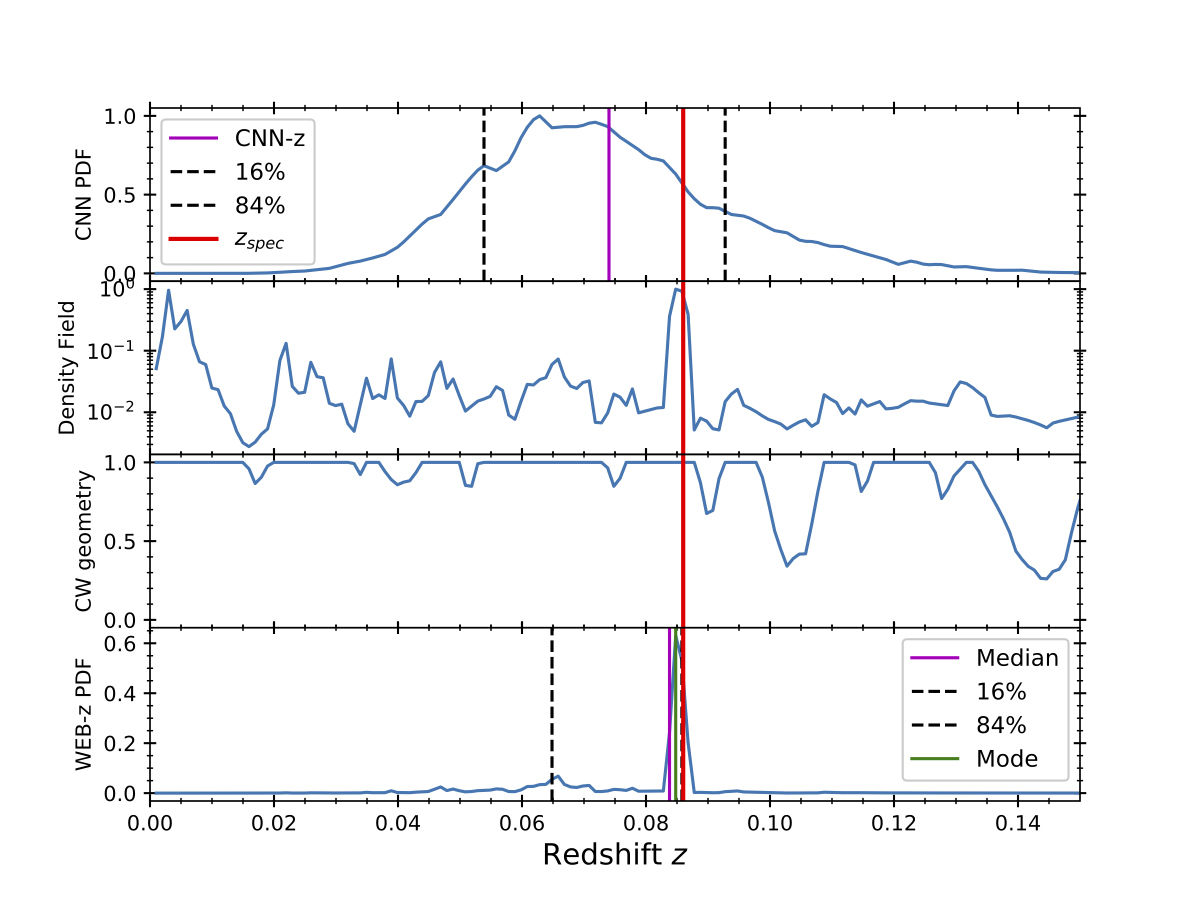}     
      \caption{ 
              }
         \label{pw-pdf-f}
    \end{subfigure}
   \begin{subfigure}[b]{0.5\textwidth}
            \includegraphics[width=0.85\hsize]{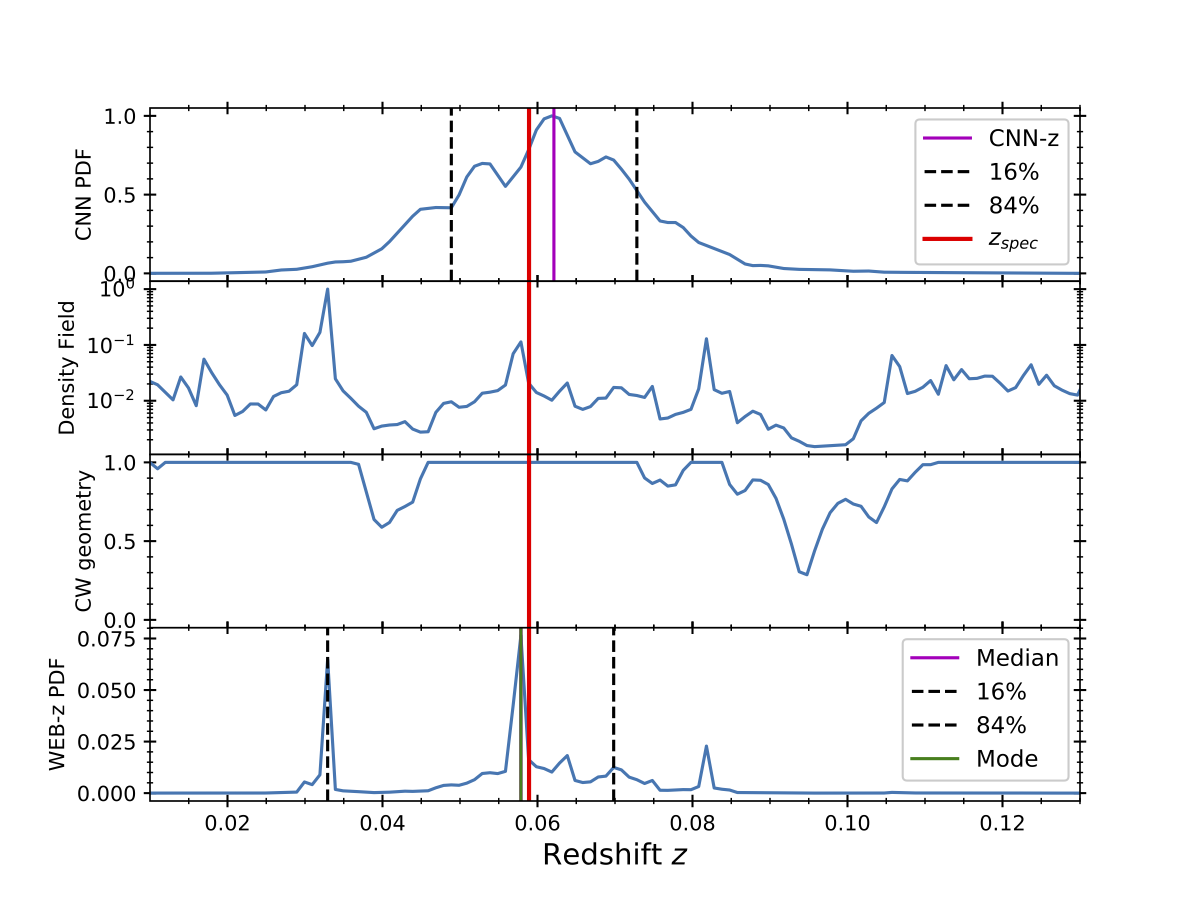}     
      \caption{ 
              }
         \label{pw-pdf-g}
   \end{subfigure}
   \caption{Random examples of PDFs obtained with the PhotoWeb method for four sources. The symbols are the same as Fig.\ref{fig:zweb}.   }
   \label{pw-pdf}
        \end{figure*}

 Additional illustrations of the \zweb method for randomly selected galaxies are presented in Fig.~\ref{pw-pdf}. 
%

\end{document}